\documentclass[%
 reprint,
superscriptaddress,
 amsmath,amssymb,
 aps,
prx,
showkeys
]{revtex4-2}
\usepackage{xcolor}
\usepackage{graphicx}% Include figure files
\usepackage{dcolumn}% Align table columns on decimal point
\usepackage{bm}% bold math
\begin{document}

%\preprint{APS/123-QED}

\title{Targeted Calibration to Adjust Stability Biases \\ in Complex Dynamical System Models}% Force line breaks with \\

\author{Daniel Pals\textsuperscript{‡}}
 \affiliation{Ludwig-Maximilians-University, Theresienstr. 37, 80333 Munich, Germany}
 \affiliation{Formerly at: Technical University of Munich, Germany; Munich Climate Center; TUM School of Engineering and Design, Department of Aerospace and Geodesy, Earth System Modelling Group}
\thanks{These two authors contributed equally.}

\author{Sebastian Bathiany\textsuperscript{‡}}
 \email{sebastian.bathiany@tum.de}
 \affiliation{Technical University of Munich, Germany; Munich Climate Center; TUM School of Engineering and Design, Department of Aerospace and Geodesy, Earth System Modelling Group}
 \affiliation{Potsdam Institute for Climate Impact Research, Telegrafenberg A 31, Potsdam, 14473, Germany}
\thanks{These two authors contributed equally.}
%\collaboration{MUSO Collaboration}%\noaffiliation

\author{Joel Kuettel}
\affiliation{Technical University of Munich, Germany; Munich Climate Center; TUM School of Engineering and Design, Department of Aerospace and Geodesy, Earth System Modelling Group}
\affiliation{Climate and Environmental Physics, Physics Institute, University of Bern, Bern, Switzerland}

\author{Richard A. Wood}
\affiliation{Met Office Hadley Centre, FitzRoy Road, Exeter, EX1 3PB, United Kingdom}

\author{Niklas Boers}
\email{n.boers@tum.de}
\affiliation{Technical University of Munich, Germany; Munich Climate Center; TUM School of Engineering and Design, Department of Aerospace and Geodesy, Earth System Modelling Group}
\affiliation{Potsdam Institute for Climate Impact Research, Telegrafenberg A 31, Potsdam, 14473, Germany}
\affiliation{Department of Mathematics and Global Systems Institute, University of Exeter, North Park Road, Exeter, EX4 4QE, UK}
%\collaboration{CLEO Collaboration}%\noaffiliation

%\noaffiliation
\renewcommand{\thefootnote}{\fnsymbol{footnote}}
%\footnotetext[1]{These authors contributed equally.}
%\renewcommand{\thefootnote}{\arabic{footnote}}

\date{\today}% It is always \today, today,
             %  but any date may be explicitly specified

\begin{abstract}
Models of complex dynamical systems like the Earth's climate often involve large numbers of uncertain parameters. Comprehensive exploration of the parameter space is typically prohibitive due to excessive computational costs, and systematic gradient-based parameter optimization is not feasible because such models are typically not differentiable. This is especially problematic in cases where the models intend to describe highly nonlinear and possibly abrupt dynamics, where sensitivity to parameter changes is high. Components of Earth's climate system, such as the North Atlantic Overturning Circulation, the polar ice sheets, or the Amazon rainforest, are at risk of undergoing critical transitions in response to anthropogenic climate change. However, estimates of the critical forcing thresholds are highly uncertain because the parameter spaces of complex climate models cannot be fully explored. Concerns have been raised that the above Earth system components are too stable in state-of-the-art models. Here, we introduce a method for efficient, systematic, and objective calibration of dynamical complex system models, targeted at adjusting system stability. Given a number of physical or observational constraints, our method moves the system in a direction where the system loses or gains stability, guided by indicators of ‘critical slowing down‘. 
In contrast to a brute force approach, where the computational cost would exponentially increase with the number of parameters, our method scales polynomially and thus evades the curse of dimensionality. We successfully apply our method to a conceptual double-fold bifurcation model and a physically plausible reduced-order model of the global ocean circulation. Our method can efficiently adjust stability biases across a range of complex system models, helping to reveal hidden instabilities and resulting state transitions they induce. In particular, it can help us to explore and improve the representation of key multistable components of the Earth system in climate models.
\end{abstract}


\maketitle

%\tableofcontents

\section{\label{sec:Intro}Introduction}

%% Complex climate models
Numerical models of complex systems such as the Earth's climate are difficult to control because they are computationally expensive to run and depend on many free or at least highly uncertain parameters. Given the nonlinear dynamics typically represented by such models, they can respond to changes in these parameters in drastic and unforeseen ways. Prominent examples are global climate models, ranging from reduced-order box models with a few coupled equations and Earth system models of intermediate complexity (EMICs, \citep{Claussen2002}) to General Circulation Models (GCMs), which explicitly simulate 3D fluid dynamics in atmosphere and oceans, and complex Earth system models (ESMs), which are used in the Coupled Model Intercomparison Project CMIP \citep{Eyring2016} and typically include biogeochemical cycles \citep{Gutierrez2021, Scholze2012}. ESM components and other geoscientific models that capture individual components of the Earth system (such as ocean-only models, terrestrial vegetation models, and ice sheet models) are often also applied in an ``offline'' (standalone) fashion and feature similar challenges.

Due to their complexity, and the limited availability of observations, such models involve unavoidable and often unquantifiable uncertainties of structural and parametric origin \citep{Curry2011, Stainforth2007}. In particular, ESMs exhibit a large number, at the order of hundreds, of free parameters, e.g. from representing sub-grid-scale processes that cannot be explicitly resolved, such as turbulent mixing, cloud formation, or biogeochemical phenomena. Even climate models of reduced complexity \citep{Claussen2002} are still too complex for a systematic assessment of their full behavior or comprehensive exploration of parameter spaces. On the other hand, low-complexity conceptual models \citep{Dijkstra2024} can lack consistency with more realistic models.

The parametric uncertainty of global climate models is particularly problematic in the context of so-called tipping points. These are critical thresholds in forcing where a component of the climate system can respond abruptly and reorganize into another state, potentially in an irreversible way \citep{Lenton2008,Chen2021,Wang2023,Boers2025}. Such catastrophic phenomena can in some cases be associated with bifurcations in the underlying dynamics \citep{Boers2022}. 
%points, i.e. parameter values where the number and stability of fixed points suddenly changes.

%% AMOC
The most prominent elements of the Earth system that have been suspected to be able to show such tipping behavior under anthropogenic forcing are the Atlantic Meridional Overturning Circulation (AMOC), the ice sheets on Greenland and Antarctica, and the Amazon rainforest \citep{Lenton2008,Boers2025,Wang2023}. However, the associated critical forcing levels remain unconstrained because transitions in any of the above systems in response to anthropogenic forcing would be unprecedented and because the uncertainties in the few climate model simulations targeted at quantifying these thresholds remain large. Current models do not show robust agreement on such events in future projections \citep{Drijfhout2015, Terpstra2025}. Moreover, there are reasons to believe that the above systems are in general too stable in state-of-the-art models compared to their real-world counterparts \citep{Valdes2011, Mecking2017}. This implies that undesired surprises, which even the most sophisticated and comprehensive ESM projections cannot warn of, could occur in the (near) future. Methods to address existing stability biases in climate models and to identify plausible worst-case scenarios are, therefore, urgently needed. 

A well-established framework for diagnosing stability changes from time series, e.g. from observations, relies on the phenomenon of critical slowing down (CSD) \citep{Haken1983, Scheffer2009, Boers2025}. When the linear stability of a stable equilibrium state is reduced and finally lost, e.g. when reaching a bifurcation, the return time to equilibrium increases, and (assuming a continuous system at a fixed point) the largest negative eigenvalue of the Jacobian for small perturbations increases toward 0 (toward 1 in the discretized representation of the system). In addition, if the system is permanently perturbed by stationary noise, the autocorrelation of its state increases toward 1 \citep{Held2004, Scheffer2009}. CSD-based indicators have been found to increase in observations of a number of suspected Earth system tipping elements such as the Greenland ice sheet \citep{BoersRypdal2021}, the AMOC \citep{Boers2021}, the Amazon rainforest \citep{Boulton2022,SmithTraxlBoers2022}, or the South American monsoon \citep{Bochow2023}; see also \cite{Boers2025} for a recent review. CSD indicators have also been demonstrated to work in high-dimensional climate models \citep{Held2004, Kleinen2003, Boulton2014}, and have been successfully used to identify the perturbation pattern to which a system is least resilient \citep{Bathiany2013a, Bathiany2013b, Weinans2019}. Although CSD-based indicators can diagnose a change in stability over time, they cannot quantify stability in an absolute sense, or quantify how far in parameter space, or even in time, the system is from tipping \citep{Ben-Yami2024}.

%%% existing methods for parameter calibration / tuning
In this study, we suggest a new CSD-based method to efficiently change the stability of a system in a given model via automatic model parameter updates, with the purpose of adjusting potential stability biases or revealing tipping risks in complex system models. Our approach resembles concepts in control theory \citep{Astroem1973, Baumeister2018}, but with the aim of altering the stability of the system, rather than keeping it in a functioning regime. Our approach is also related to inverse problems, which seek parameter values that make the model’s output more consistent with observations. In our context, however, we aim to identify parameter values that only change the system's stability (a property that cannot be directly observed), while keeping the observables constant.

Recently, atmospheric models that are purely based on machine learning \citep{Price2025, Bi2023, Watt-Meyer2025} and a differentiable hybrid model \citep{Kochkov2024} have been developed, but such models struggle to show a physically realistic response to forcing \citep{Rackow2024}. More traditional climate models are more physics-based but are not differentiable, i.e. gradients of model output with respect to parameter changes cannot be directly computed, which hinders systematic and objective parameter optimization \citep{Gelbrecht2023, Shen2023}. Therefore, these models have typically been hand-tuned based on subjective expert judgment, with the aim of approximately matching observed features, such as the global radiative balance or a target climate sensitivity \citep{Mauritsen2012, Hourdin2017, Mauritsen2020}. Although we generally strongly advocate making climate models differentiable \citep{Gelbrecht2023} to enable efficient and transparent parameter optimization with respect to observations, differentiability will be less helpful in adjusting the stability of systems in a given model. We are interested in a targeted dynamical calibration, modifying parameters that specifically affect the linear stability of an equilibrium state while keeping observables and emergent properties unchanged. This could not easily be achieved by differentiable programming or by emulating the numerical model in question with a differentiable machine learning model. In the adjoint-based optimization of differentiable models, where stability metrics would be differentiated through long, potentially chaotic integrations, gradients become ill-conditioned when computed over chaotic dynamics \citep{Lea2000}, and especially near bifurcations where relaxation times increase dramatically. Our approach avoids many of the problems related to ill-conditioned gradients of either natively differentiable models or (differentiable) machine learning emulators of non-differentiable models. 

Derivative-free calibration methods like simplex methods \citep{Nelder1965} or Ensemble Kalman Inversion \citep{Kovachki2019} can also solve inverse problems by generating a perturbed-physics ensemble \citep{Collins2011}, comparing the output to desired (typically observed) values, and then iterating the ensemble of model parameters. This strategy can involve so-called emulators, which aim to capture the model's parameter-to-output relationship, but can be evaluated (``sampled'') in a much cheaper way \citep{Cleary2021, Yang2025}. Related Bayesian approaches even aim to provide an uncertainty distribution on the parameters in question \citep{Cranmer2020, Lux2022}. A different line of work is to train a machine learning emulator on a sparse number of parameters and construct the bifurcation diagram by solving the characteristic polynomial, leveraging the differentiability of the trained emulator \citep{Kato2025}. However, running even small ensemble simulations can still be computationally costly. Moreover, emulation may be challenging for very nonlinear features like tipping events, and in situations when one needs to extrapolate into unexplored regions of parameter space. Even if such approaches were technically successful, a probabilistic interpretation of parameter values or model outcomes regarding tipping points in the real world is out of reach due to the structural uncertainty of climate models \citep{Curry2011, Stainforth2007}. The problem of targeted stability tuning is complementary to the emulator approaches mentioned above. The task at hand is to efficiently identify unstable regions in the model's phase space. Current emulation approaches such as Gaussian Processes would, relying on reasonably smooth functions, not be suitable in this context. Moreover, Latin Hypercube sampling (alone) would sample the relevant parameter regions too sparsely.

%% our method
Here, we present a CSD-based method to systematically and objectively adjust the stability of a given state of a simulated system. The method works sequentially, without the need for costly ensemble simulations. The main types of models that we target here are relatively fast ones, allowing simulations that span several thousand times the characteristic timescale of the nonlinear system in question. Also, our method tends to work best when the model does not feature internal (chaotic) variability. This potentially includes any Earth system model component except general circulation models (GCMs), Earth system models of intermediate complexity (EMICs), lower complexity models that serve as emulators (simplified versions) of complex ESMs (e.g. box models, conceptual models, or data-driven machine-learning models). Ideally, these models are dynamical system simulators that share properties with state-of-the-art ESMs, e.g. box models calibrated to specific target ESMs \citep{Wood2019, Chapman2024a, Chapman2024b}, or subcomponents taken out of more comprehensive models \citep{Bathiany2024}, which will allow conclusions even for comprehensive ESMs.

Our method considers dynamics on the combination of a model's phase and parameter space. As described in detail in the following sections, we create a feedback loop between the local stability of the system's state and its parameter values. Based on a certain initial state of the system and its parameters, our method determines the direction in parameter space where the system most effectively loses (or gains) stability under certain observational or physical constraints. Our general approach to this problem is to find a local parameter-dependent autoregressive model which effectively describes the dynamics of an observable close to the stable state of interest, and then use this model to find a new parameter combination which increases or decreases the stability of the given equilibrium state. Using the CSD phenomenon makes our approach for targeted and objective model (re-)calibration highly efficient; as we will show below, our method scales polynomially with the numbers of parameters, whereas a brute-force perturbed-parameter ensemble approach would scale exponentially (the ``curse of dimensionality''). 

We describe our method in its most general way in the following section, and then apply it to two example systems featuring multistability, demonstrate its efficiency, and discuss the implications of our results. Our study is accompanied by public code that can also be applied to other systems (see below). 

\section{Targeted model calibration to adjust system stability}
\label{sec:Recipe}

We consider a general dynamical system that is discrete in time, which essentially covers all numerical models of dynamical systems, including climate and Earth system models. We denote the dynamic variables (state variables) of the system by $\vec x \in \mathbb R^{d_x}$ and the parameters by $\vec p \in \mathbb R^{d_p}$. Moreover, we consider observables $\vec o \in \mathbb R^{d_o}$. We assume that we have access to a function $f_o: \mathbb{R}^{d_x + d_p} \rightarrow \mathbb{R}^{d_o}, \; (\vec x, \vec p) \mapsto \vec o$ mapping the state variables to the parameter-dependent observables $(\vec o_t)$. As an example, the system could be the global ocean circulation, and the observable could be the mass flux across a certain latitude in the North Atlantic (AMOC strength).

The dynamics of the system is given in the form of an evolution function 
$$
f_p: \mathbb{R}^{d_x + d_p} \rightarrow \mathbb{R}^{d_x}, \; (\vec x_t, \vec p) \mapsto \vec x_{t+1}\,,
$$ 
which defines the one-step ahead propagation of the system state $\vec x_t$ to $\vec x_{t+1}$ and allows us to integrate it. It is not necessary to know the exact expression of $f_p$; we only need the ability to run parameter-dependent simulations of the system. We assume that the system's state fluctuates around a dynamic equilibrium which has a certain local stability, defined in terms of the linear restoring rate. The target is to adjust the system's parameters in a way that changes this local stability.

\begin{figure}
\centering
\includegraphics[width=.8\linewidth]{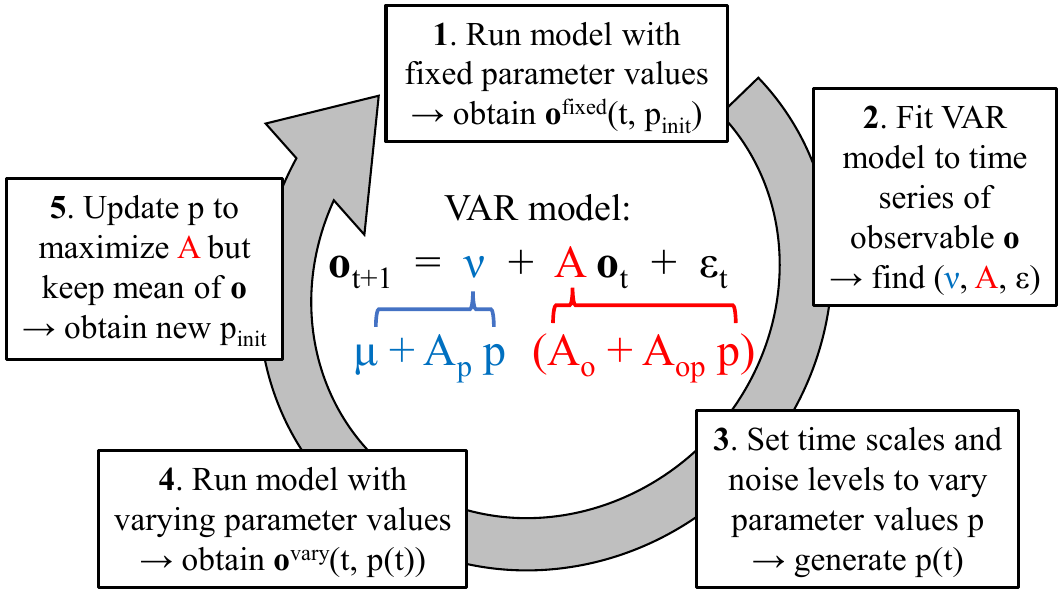}
        \caption{Main steps of our calibration method to automatically update parameter values that change the stability of a simulated system in a targeted way. o are observables constructed from the output of the complex model, p are the model parameters to be calibrated (in general, both are vectors). A vector-autoregressive model (VAR) is fitted to the data with the aim of varying p such that it maximizes the largest eigenvalue of matrix A (red), while keeping the time mean of o constant. The nature of vector and matrix multiplications are not specified here; for more detail on each step see Sect. \ref{sec:Recipe}.}
         \label{fig: method}
\end{figure}

Our method consists of the following iterative steps (Fig. \ref{fig: method}), which all run fully automatically once a system has been set up. For the sake of clarity, we focus on the case of destabilizing a given system; adjustments to the case of increasing stability are straightforward. 

%% naming convention for noise and error terms:
%  parameter noise is w, and state noise is called u, with superscript "fixed" or "vary". Error terms of VAR models are called epsilon, again with "fixed" and "vary".

\begin{itemize}
    \item[1.] Generate a trajectory of observables $(\vec o_t)$ of length $T$ with parameters fixed to their initial values $\vec p_{init}$. To this end, we integrate the system to obtain a trajectory of the system variables $(\vec x_t)$, to which we apply $f_o$. During the integration process we force the underlying system with small additive white noise $\vec u_t^{fixed} \in \mathbb R^{d_x}$ (``fixed'' because the parameters are fixed here) in order to drive the system out of equilibrium and control the scale of the region in phase space used to compute the Jacobian at equilibrium. 
    \begin{align}
        \vec x_{t+1} =& f_p(\vec x_t, \vec p_{init}) + \vec u_t^{fixed} %\nonumber 
        \\
        \vec o_{t} =& f_o(\vec x_t, \vec p_{init}) %\nonumber 
    \end{align}
    It generally suffices to use state noise levels that are so small that they do not affect the system's general state and its dynamics. The only purpose is to be able to sample the system's linear stability around a fixed point.
    
    \item[2.] Fit a vector-autoregressive model (VAR(1) model) of the form 
    \begin{align}%\nonumber 
    \label{eq:VARfixed}
        \vec o_{t+1} = \vec \nu + A \vec o_t + \vec \varepsilon_t^{fixed}
    \end{align}
    to the data from step 1, including uncertainty ranges (see Appendix \ref{sec: step2_VARmodel} for theoretical context on this approach), where $\vec \nu$ is a constant baseline state, the square matrix $A$ contains constant parameters of the auto-regressive model, and $\vec \varepsilon_t^{fixed}$ are the residuals of the fit (resulting both from the imposed noise and the potential model error). Since the equilibrium is stable, all eigenvalues of $A$ should be smaller than one and the stability of the system is characterized by the largest eigenvalue $\lambda_{max}$.
    %For this work we assume that all dynamical system we consider model some continuous time processes. This implies that bifurcations of these models correspond to the event where some eigenvalue of the Jacobian, evaluated at the considered fixed point of the discretized system, passes 1 from below, changing the stability of the fixed point. I.e. we in particular assume all eigenvalues to be greater than zero. 
    
    \item[3.] Vary the parameter values. To this end, we use univariate autoregressive (AR(1)) processes of the form
    \begin{align}\label{eq: parameter variation}
        p_{i,t+1} = (1-\lambda_p) \, p_{i,init} + \lambda_p \, p_{i,t} + \sigma_i w_{i,t}
    \end{align}
    to generate a time series for each parameter $p_i$, fluctuating around $p_{i,init}$, where $(\vec w_{i,t})$ describes white noise with $w_{i,t} \sim \mathcal N(0,1)$.  Here, we choose the persistence parameter of the AR(1) process, $\lambda_p$, such that $\lambda_p \leq \lambda_{max}$. In this way, the parameters vary on a timescale that is longer than the timescale of the system, expressed as the slowest mode of the observable dynamics. This is necessary in order to record the response of the system to the parameter changes. We set the scalar $\lambda_p$ to $\lambda_{max}$ plus two standard deviations from $\lambda_{max}$ to ensure this condition. In Appendix \ref{sec: step3_noiseamplitudes}, we discuss how to find appropriate noise amplitudes $\sigma_i$ and in Sect. \ref{sec: application} we discuss how we implemented the method based on the model at hand.
    
    \item[4.] Further integrate the system, forcing it with the time-dependent parameter values: 
    \begin{align}
        \vec x_{t+1}^{vary} =& f_p(\vec x_t^{vary}, \vec p_{t}) + \vec u_t^{vary} %\nonumber 
        \\
        \vec o_{t}^{vary} =& f_o(\vec x_t^{vary}, \vec p_{t}) %\nonumber 
    \end{align}
    
    Our aim is to use these simulations to understand how parameter values influence the stability - and consequently the VAR parameters - of the system.
    
    Our approach is to fit a VAR(1) model of the form
    \begin{eqnarray}
        \label{eq:VARvary}
        \vec o_{t+1}^{vary} &=& \vec \mu + A_o \vec o_t^{vary} + A_p \vec p_t \nonumber
    \\
    &&+ A_{op} (\vec o_t^{vary} \otimes \vec p_t) + \vec \varepsilon_t^{vary}
    \end{eqnarray}
    with $A_o \in \mathbb R^{d_o \times d_o}$, $A_p \in \mathbb R^{d_o \times d_p}$, $A_{op} \in \mathbb R^{d_o \times (d_o \cdot d_p)}$ and the Kronecker product
    \begin{eqnarray} %\nonumber 
        \vec o \otimes \vec p &\equiv& (o_1 p_1, o_1 p_2, ..., o_1 p_{d_p}, ..., o_{d_o} p_1,..., \nonumber \\ && o_{d_o} p_{d_p})^T \in \mathbb R^{d_o \cdot d_p}\,, 
    \label{eq:Kronecker_product}
    \end{eqnarray}
    where $d_o$ is the number of observables and $d_p$ the number of parameters.

In the above, the second-order terms (mixed terms involving $\vec o$ and $\vec p$) are needed to detect the p-dependence of the eigenvalues (represented by matrix $A$ in Eq. \ref{eq:VARfixed}). We note that the quantification of these terms can become highly uncertain if the first order effects are much more prominent than the second order effects. In this case, it would be possible to shift contributions between $\vec \mu$, $A_o$, $A_p$ and $A_{op}$ in Eq. \ref{eq:VARvary}, i.e., we would be faced with an underdetermined problem. We solve this potential problem by exploiting our knowledge of the state noise (see Appendix \ref{sec: step4_secondorderterms}). 

\item[5.] Based on the estimated model from Eq. \ref{eq:VARvary}, we choose new parameter values. Our goal here is to maximize the largest eigenvalue of $A_o + A_{op} (\mathbb{I}_{d_o} \otimes \vec p)$ (where we used Eq. \ref{eq:Kronecker}, and where $ \mathbb{I}_{d_o}$ is the Identity matrix of size $d_o \times d_o$ and $\otimes$ is the Kronecker product is defined in Eq. \ref{eq:Kronecker_product}). We do not allow the eigenvalue to exceed 1, as we only aim at reducing the stability of the fixed point, and do not intend to make it fully unstable. Other calibration targets are possible in different contexts.
    
We also demand that the updated parameter values should not differ too much from the previous ones, as the VAR model is only valid in a vicinity of these values. In special cases, more efficient solutions are possible (e.g. the double-well potential in Sect. \ref{sec: Double well system}). In addition, we enforce a preservation of the mean values of the observables, which turns Eq. \ref{eq:VARvary} into
\begin{align}\label{eq: preserving bar o}
        \bar o \overset{!}{=} \vec \mu + A_o \bar o + A_p \vec p + A_{op} (\bar o \otimes \vec p)
\end{align}
with $\bar o$ denoting the initial equilibrium values of the observables. This condition is motivated by the fact that the mean state (for example of the climate system) is typically constrained by observations much better than model parameters or the stability of the state. Appendix \ref{sec: step5_parameterupdates} provides details on how we use the above conditions to determine parameter updates. Model-specific details in applications of our method are discussed in the following section.

\item[6.] After finding a new set of parameters, we replace $\vec p_{init}$ by the new parameter values and run the system until it equilibrates. After that, the whole process is repeated, now using the new parameter values.
\end{itemize}

\section{Application to two example systems}
\label{sec: application}
In order to test our method, we apply it to a simple double-well dynamical system, as well as to a conceptual, physically plausible model of the global ocean circulation \citep{Wood2019} (see Appendix \ref{sec: five-box model description}).

\subsection{Double-well system}\label{sec: Double well system}

We consider a simple double-well system with dynamics determined by the ODE
\begin{align}
    \label{eq: Double well ODE}
    \dot x = p_1 x^3 + p_2 x^2 + p_3 x + p_4
\end{align}
For simplicity, we choose the observable $o$ to be $x$ itself, i.e. $f_o(x,\vec p) = x$. We use the following initial parameter values:
\begin{align}%\nonumber
    \vec p_{init} = \begin{pmatrix}
        p_1 \\
        p_2\\
        p_3\\
        p_4
    \end{pmatrix}
    = \begin{pmatrix}
        -1\\
        0\\1\\0
    \end{pmatrix}
\end{align}

With this choice the system has two stable fixed points at $x = 1$ and $x = -1$ and one unstable fixed point at $x = 0$. Each fixed point has its own basin of attraction and the two basins are separated by the unstable fixed point. We aim to destabilize the fixed point at $x = 1$ with regard to small perturbations in the $x$-direction whilst maintaining the position of the fixed point at $x = 1$.

As state noise $u_t$, we add Gaussian white noise with a noise level of $10^{-4}$ to the right-hand side of Eq. \ref{eq: Double well ODE}. We integrate the resulting stochastic differential equation using the Euler-Maruyama scheme \citep{Kloeden1992}, where we choose a time step of $\Delta t = 10^{-3}$.

We run the destabilization process for 18 iterations since further iteration would typically lead to a stochastic escape into the alternative basin of attraction. The noise amplitudes used for the parameter variation are updated in each iteration step (see Sect. \ref{sec:Recipe}, step 3, and Appendix \ref{sec: step3_noiseamplitudes}). We calibrate these by first setting all parameter noise amplitudes $\sigma_i$ simultaneously to $10^{-3} \cdot \sqrt{1-\lambda_p^2}$ and using these to compute varying parameter trajectories via Eq. \ref{eq: parameter variation}. The system, forced by the varying parameters, is then integrated for 100 time steps (this number must be chosen large enough to allow the distribution of the observable to adjust to the new parameter noise). 
This is repeated iteratively whilst increasing the noise amplitudes in each iteration by a factor of 2, up to the point where the standard deviation of $x$ exceeds a value of $10^{-3}$. The noise amplitudes for generating the parameter series are then set to the values prior to the termination condition. To ensure that the VAR model actually shows a dependence on the parameter variations, we check whether the coefficients $A_p$ and $A_{op}$ of the parameter-dependent VAR model differ from zero (using the estimated errors of the coefficients, see Appendices \ref{sec: step2_VARmodel} and \ref{sec: step3_noiseamplitudes}). This turns out to always be the case. With this method we typically find noise amplitudes corresponding to a standard deviation of the parameters (given by $\sigma_i/\sqrt{1-\lambda_p^2}$ for parameter $p_i$) greater or close to 0.05.

We estimate the parameter-dependent VAR model using 1000 time steps for the fixed parameter model and 100 time steps for the parameter dependent part. The parameters are then updated at the end of each iteration step by maximizing $A_o + A_{op} \vec p$, which is a scalar in this system and hence already the eigenvalue we want to increase via the parameter change. The constraint of preserving the mean value $\bar x = 1$ (see Appendix \ref{sec: step5_parameterupdates} and Eq. \ref{eq: preserving bar o}) then reads 
\begin{align}%\nonumber
    1 \overset{!}{=} \mu + A_o + (A_p + A_{op} ) \vec p\,.
\label{eq: constraint_DWS}
\end{align}
The system can now be destabilized under constraints as described in more detail in Appendix \ref{sec: step5_parameterupdates}. 

Since the system is relatively simple, an even more efficient variant also works here. The difference to the more general approach described in Appendix \ref{sec: step5_parameterupdates} is that the constraint can be strictly imposed here instead of using a loss function. To this end, we choose a parameter update $\Delta p$ of the form $\Delta p =  \Delta \vec p_{\bot} + L \Delta \vec p_{\|}$ with
\begin{align}
    \Delta \vec p_{\bot} =& \frac{1 - \mu - A_o - \vec n \cdot \vec p_{init}}{|\vec n|^2} \, \vec n %\nonumber
    \label{eq: p_perpendicular}
    \\
    \Delta \vec p_{\|} =& \left[ A_{op} - \frac{A_{op} \cdot \vec n}{|\vec n|^2} \vec n \right]\,, %\nonumber
    \label{eq: p_parallel}
\end{align}
where $L \in \mathbb R$. 

Note that in the double-well system, $A_o$ and $\mu$ are scalars, and $A_{op}$, $A_p$, $\vec p$ and $n$ are all vectors of dimension 4, where $n = A_p + A_{op}$ is the normal vector to the hyperplane $\mathcal{H}$ spanned by all possible $\vec p$ that satisfy Eq. \ref{eq: constraint_DWS}. To be consistent, we here only show arrows above vectors that are vectors in any application, not only the double-well system. Eq. \ref{eq: p_perpendicular} and \ref{eq: p_parallel} are the solution to the optimization problem where $A_o + A_{op} \cdot \vec p$ (eigenvalue) must be maximized, and $(A_p + A_{op}) \cdot \vec p$ must remain constant (constraint $\bar x = 1$). 

$\Delta \vec p_{\bot}$ is the component perpendicular to the hyperplane $\mathcal{H}$ (and parallel to $n$). By adding $\Delta \vec p_{\bot}$ to $\vec p_{init}$, we move $\vec p$ onto $\mathcal{H}$ in an orthogonal manner, in order to always keep $\bar x$ at 1. $\Delta \vec p_{\|}$ is the component parallel to $\mathcal{H}$. Adding a multiple of $\Delta \vec p_{\|}$ to our parameter vector thus destabilizes the system without leaving $\mathcal{H}$. 

By then choosing
\begin{eqnarray}
\label{eq:doubleWellLambda}
    L &=& \mathrm{min}\Big(\frac{0.05}{|\Delta p_{\|,1}|}, ..., \frac{0.05}{|\Delta p_{\|,4}|}, \nonumber \\ && \frac{1 - A_o - A_{op}( \vec p_{init} + \Delta \vec p_{\bot})}{2 \cdot A_{op} \vec p_{\|}}\Big)
\end{eqnarray}
we assure by the first four arguments of the min(...) function -- assuming that $|\Delta \vec p_{\bot}|$ is comparably small -- that the parameter changes are not too large compared to the region explored in parameter space. The last argument of the min(...) function preserves the stability of the equilibrium at $x = 1$ by not pushing the value of $A_o + A_{op} \vec p$ past 1. To this end, we compute the value that $L$ would have to take in order for $A_o + A_{op} (\vec p_{init} + \Delta \vec p)$ to be equal to one and use this value, divided by a factor of 2.

\begin{figure}
         \centering
         \includegraphics[width=\linewidth]{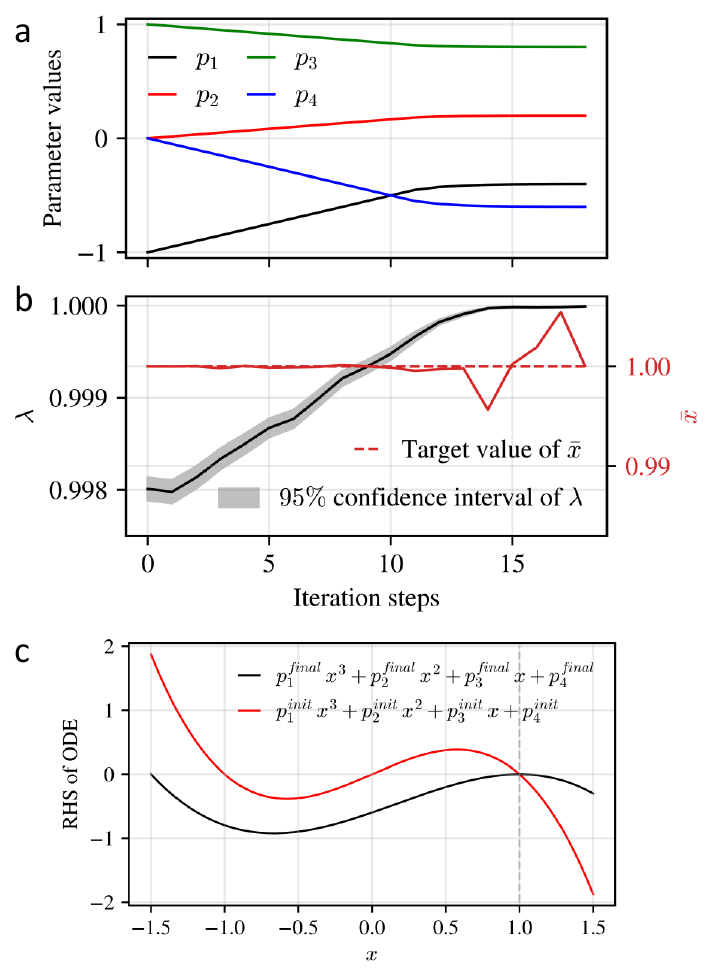}
        \caption{Application of the destabilization method to the double-well system. a) Evolution of the parameters $p_1$,...,$p_4$, and (b) evolution of the Jacobian $\lambda$ and the mean value $\bar x$ of $x$ during the iterative destabilization of the double-well system; note the small y-axis range for the latter.
        c) The right-hand side (RHS) of Eq. \ref{eq: Double well ODE} for the parameters before (red) and after (black) the destabilization. By design, observable $x$ is constrained to stay at $x=1$ (grey dotted line). The original stable state (negative slope of the red line) merges with an unstable one in a saddle-node bifurcation when the system is fully destabilized.}
         \label{fig: double well}
\end{figure}

The evolution of the updated parameters at the end of each iteration step is shown in Fig. \ref{fig: double well}a together with the evolution of the Jacobian (eigenvalue) $\lambda$, which indicates the stability of the equilibrium. Here, $\lambda$ moving closer to 1 from below corresponds to a stability reduction of the system. The average value $\bar x$ of $x$ computed from a time series (obtained by integrating the stochastically forced system) consisting of $10^5$ data points, using the parameter setting of the respective iteration step, can be successfully stabilized by our method (Fig. \ref{fig: double well})b. 

We notice that for the final iteration steps, both $\lambda$ and the parameter values start to converge to prevent the equilibrium from losing its stability entirely. This is an optional feature we explicitly implemented into the parameter update scheme by adding the last argument to the min(...) function in Eq. \ref{eq:doubleWellLambda}.
Another striking aspect of Fig. \ref{fig: double well} is that although the average value $\bar x$ of $x$ remains very close to its initial value in each iteration, there is a noticeable increase in fluctuations away from this value as $\lambda$ approaches 1. The reason for this phenomenon likely lies in the fact that we did not change the amplitude of the noise $u_t^{(x)}$ added to the system; so as the stability of the fixed point decreases, the standard deviation of $x$ and therefore also of $\bar x$ increases \citep{Scheffer2009}.

The effect of the destabilization process on the RHS of Eq. \ref{eq: Double well ODE} is visualized in Fig. \ref{fig: double well}c. The equilibrium at $x=1$ is significantly destabilized while maintaining its original position. We note that the parameter change induced by our calibration method has shifted the other stable equilibrium, initially located at $x = -1$, to smaller values of $x$, while simultaneously increasing the size of the basin of attraction of that fixed point.

\subsection{AMOC five-box model}\label{sec: AMOC 5-box model}
\begin{figure} %[\sidecaptionrelwidth][t!]
\centering
\includegraphics[width=\linewidth]{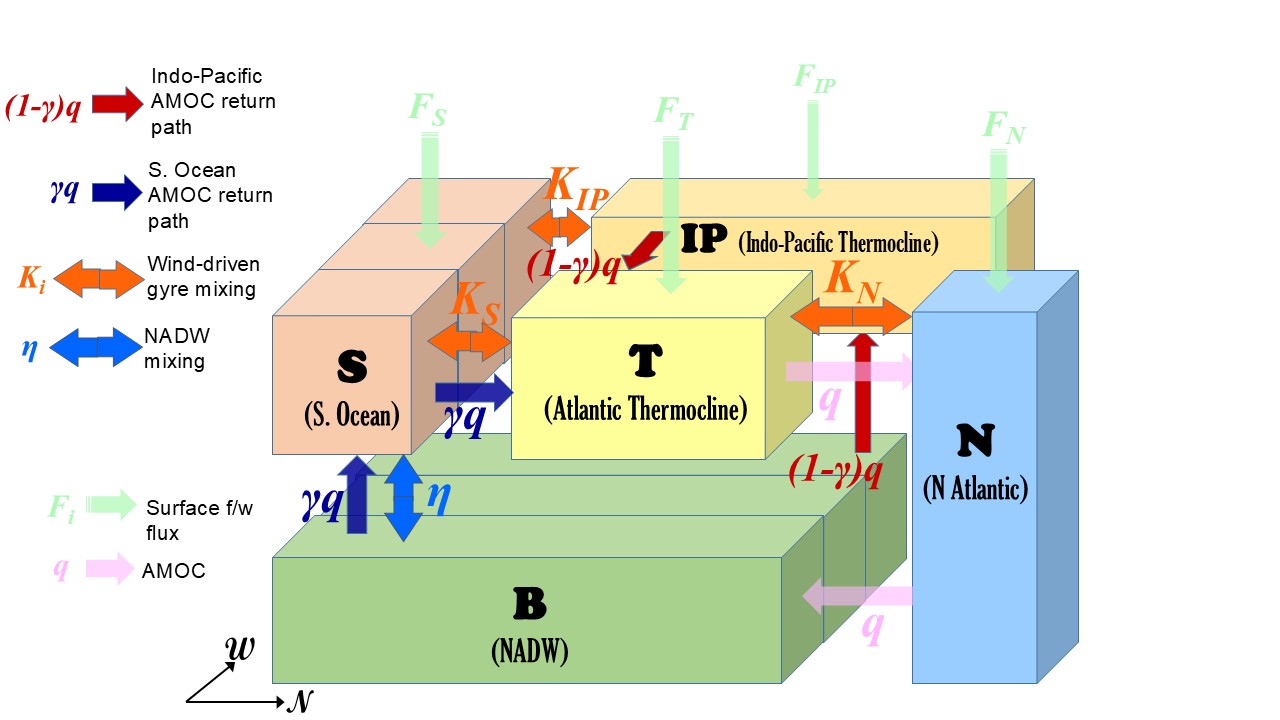}
\caption{Schematic illustration of the five-box ocean model (adapted from \cite{Wood2019}). The five boxes represent major ocean basins, and arrows represent freshwater fluxes between them. The variable q (pink arrows) is the property we consider as the 'observable' in this study. Freshwater hosing is applied by modifying the green surface fluxes $F_i$. The five parameters shown in strong colours ($\gamma$, $\eta$, $K_N$, $K_S$, and $K_{IP}$) are less well constrained by observations and are the parameters we calibrate.}
\label{fig: 5-boxmodel}
\end{figure}

In order to test our method on a more complex process-based system, we apply it to a recently proposed five-box model of the overturning circulation of the global oceans \citep{Wood2019}. The model consists of five coupled differential equations describing the dynamics of the salinities in five boxes, where each box represents a water mass prevailing in a specific region of the Earth's oceans (Fig. \ref{fig: 5-boxmodel}). We made some slight modifications to the model regarding the conservation of total water mass (see Appendix \ref{sec: five-box model description}), which only have a negligible influence on the dynamics of the system. Here, the salinities $\vec S = (S_N, S_T, S_S, S_{IP}, S_B)^T$ are the system variables called $\vec x$ in the previous section.

We select the AMOC strength $q$ as our observable $\vec o$. We vary the parameters $\gamma$ (the relative strength of a circulation branch involving the Southern Ocean), $\eta$ (representing mixing of North Atlantic deep water with fresher waters), $K_N$, $K_S$ and $K_{IP}$ (representing diffusive fluxes associated with the gyre strengths in the North Atlantic, Southern Ocean, and Indo-Pacific Ocean), as these five parameters have the highest uncertainties. The other parameters of the model are kept fixed at their original values. 

% state noise, 5-box model
We discretize the system with $\Delta t = 0.1 \mathrm{yr}$. The initial parameter values are chosen according to the $\mathrm{FAMOUS}_A$ model as stated by \cite{Wood2019}. We choose the additive noise driving the system so that the overall amount of salt in the system is conserved. To this end, we add a noise vector of the form
\begin{align}%\nonumber
\vec u_t^{(x)} = \begin{pmatrix}
    u_t^{(1)}/V_N\\
    u_t^{(2)}/V_T\\
    u_t^{(3)}/V_S\\
    u_t^{(4)}/V_{IP}\\  -(u_t^{(1)}+u_t^{(2)}+u_t^{(3)}+u_t^{(4)}) / V_B
\end{pmatrix}
\end{align} 
to the dynamic equations updating the salinity concentrations $\vec S_t$, where each $u_t^{(i)}$ is a different realization of Gaussian white noise with noise level $10^{-4}$. We then run the model for 500 iterations using $10^6$ time steps (i.e. $10^5$ years) in each iteration, for estimating the VAR model with fixed parameters (method step 2 above) as well as for finding the parameter dependencies of the model (step 4) in each iteration. Also, before starting the VAR estimation, we integrate the model for a transient time of $10^5$ time steps ($10,000$ years) in each iteration to allow the model to equilibrate with the new parameter values. Due to their physical interpretation, the parameters are restricted to positive values and $\gamma$ must additionally fulfill $\gamma \leq 1$ (although it turns out that imposing this condition is not necessary since the destabilization process decreases $\gamma$).

When adjusting the noise amplitudes $\sigma_i$, which are needed to generate the parameter series (Eq. \ref{eq: parameter variation}), we calibrate each parameter separately (see Appendix \ref{sec: step3_noiseamplitudes}). This helps determine the influence of any single parameter on the system. To this end, we fix all parameters, except one, to their initial values and iteratively increase the noise amplitude $\sigma_i$ of the parameter of interest by powers of 2, starting from $\sigma_i = \mathrm{min}(10^{-5}, p_{init,i}/4 \cdot \sqrt{1 - \lambda_p^2}$). The minimum function and factor $p_{init,i}/4$ allow the variations to also be even smaller than $10^{-5}$ should any component of $\vec p_{init}$ be close to zero. For each parameter noise amplitude, we integrate the system for $10^{5}$ time steps while forcing it only by the variations of $p_i$. As outlined in Appendix \ref{sec: step3_noiseamplitudes}, the condition for stopping the iteration is when the observable's variance exceeds a threshold (here $\mathrm{VAR}(q)<0.01$), while the VAR model parameters differ from 0. Of all noise amplitudes that meet these conditions, we select the combination of noise amplitudes (one for each parameter) for which the corresponding values of $\mathrm{VAR}(q)$ do not differ by more than a factor of 4 from each other. This guarantees that the variances of the observables lie on comparable scales for each separate parameter variation. At the same time, the individual noise amplitudes are maximized to maximize the region explored in parameter space. To keep computational costs minimal, we only update the noise amplitudes every 10 iteration steps. This procedure yields noise levels in the range $\sigma_i /\sqrt{1 - \lambda_p^2} \in [5,10]$.

\begin{figure*} %[\sidecaptionrelwidth][t!]
    \centering
\includegraphics[width=\linewidth]{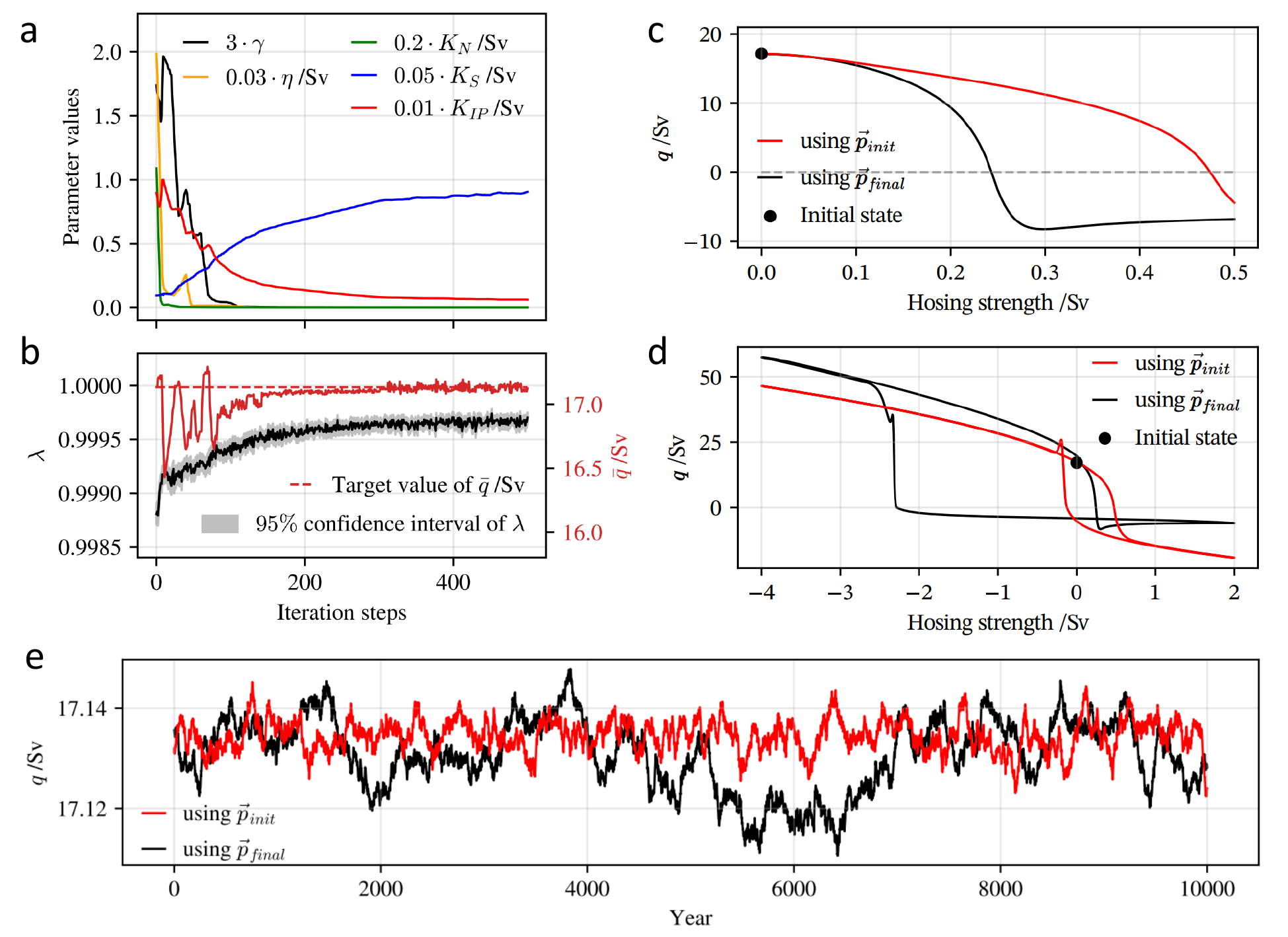}
\caption{Destabilization of the five-box AMOC model with (a) the evolution of the parameters $p_1$,...,$p_5$, as well as (b) the Jacobian $\lambda$ and the mean value $\bar x$ of $x$ during the iterative destabilization process. Increasing $\lambda$ toward 1 implies a successful destabilization of the system. (c) and (d): Hysteresis in the five-box AMOC model \citep{Wood2019} before (red) and after (black) destabilization, when performing the hosing experiment as described in \cite{Wood2019} to both systems. (c) Closeup of the hysteresis emphasizing the initial AMOC collapses due to the hosing. (d) Hysteresis curve over the full range of hosing. Note that the horizontal range of hysteresis is strongly increased by the parameter change. (e) 10,000 years of stationary time series when running the model with each parameter set, and without hosing (H=0).}
\label{fig: 5-boxmodel_destabilization}
\end{figure*}

After applying steps 1 to 5 from Sect. \ref{sec:Recipe} to find a model of the form
\begin{align}%\nonumber
\label{eq: VARmodel_5box}
    q_{t+1} = \mu + A_o q_t + A_p \vec p + q_t A_{op}  \vec p,
\end{align}
we now systematically reduce the stability of the strong AMOC state in the five-box model under a number of physical constraints, most importantly that $\Delta \vec p$ always stays on the same hyperplane. The implementation of these constraints is explained in Appendix \ref{sec: step5_parameterupdates}.

We here set the length $l$ of the parameter update vector $\Delta \vec p$ to 100 and use
\begin{align} %\nonumber 
    l_i = \mathrm{min}\left( \frac{\sigma_i}{\sqrt{1-\lambda_p^2}}, \frac{p_{init,i}}{100} \right) 
\label{eq:metric_l_AMOC}
\end{align}
with $\sigma_i$ and $\lambda_p$ from Eq. \ref{eq: parameter variation}. With this choice, we guarantee that no parameter value becomes negative and that we stay sufficiently close to the region explored in parameter space. In more general cases, the limitation to positive parameters can be dropped by setting $l_i$ to the first term in the $min()$ function (Eq. \ref{eq:metric_l}).
We set the penalty coefficients in the optimization (see Eq. \ref{eq:maximise}) to $\eta_1 = 10^{12}$, $\eta_2 = 10^3$ and $\eta_3 = 1$. The very high penalty value $\eta_1$ ensures that the mean state $\bar q$ is well conserved.

%% 5-box model results
%Analogous results as to those shown in Fig. \ref{fig: parameter evolution double well}, but for the AMOC five-box model as introduced by \cite{Wood2019}, are shown in Fig. \ref{fig: 5-boxmodel_destabilization}. 
The parameter values and the observable state converge and are close to their final values after a few hundred iterations (Fig. \ref{fig: 5-boxmodel_destabilization}a,b). From Fig. \ref{fig: 5-boxmodel_destabilization}a we can also infer that the two model parameters $K_N$ and $\eta$ have a major influence on the stability of the system, as they both rapidly tend towards zero within the first few iteration steps whilst simultaneously causing a significant decrease in the stability of the system. In principle, this result confirms that our method is physically meaningful; for example, the gyre exchange coefficients $K_i$ represent the sensitivity to wind stress, which is known to be an important factor for AMOC stability in ESMs \citep{Jackson2017}.

These parameter values, however, may be physically implausible without further constraints. For example, fresh water exchange by the wind-driven gyre circulation, represented by $K_N$, is known to be a stabilizing factor, and observations and climate models show that this process is active in the North Atlantic \citep{McDonagh2015, Jackson2017}, suggesting that the minimum allowable value of $K_N$ should be set somewhat larger than zero. In addition, we have used only a single constraint on the destabilized equilibrium solution (the overturning strength q) for simplicity. We find that our destabilized equilibrium solution has box-mean salinities that are inconsistent with observations. In particular, the box representing the Indo-Pacific oceans is too salty, while the boxes representing the North Atlantic and Southern Ocean are too fresh. Such additional observational constraints can in principle be added to our method, and would presumably result in a solution that is somewhat less unstable than the one we find here.

After destabilizing the five-box model, we verify that its sensitivity to parameter changes in the form of freshwater forcing (so-called hosing) has increased. To this end, we apply freshwater hosing as described in \cite{Wood2019} to both the original and the destabilized system (Fig. \ref{fig: 5-boxmodel_destabilization}c-d). The effect of the parameter change induced by our method is similar to the one it had on the double well system: When exposed to the same hosing, the destabilized model (black lines in Fig. \ref{fig: 5-boxmodel_destabilization}c-d) reaches an alternative steady state at much smaller hosing than the model with the original parameter values (red lines). Moreover, the negative hosing (reversed forcing) that is required for the system to recover to its original state is much larger for the destabilized system, i.e. the regime with hysteresis has become wider (Fig. \ref{fig: 5-boxmodel_destabilization}d). Moreover, Fig. \ref{fig: 5-boxmodel_destabilization}e shows that the AMOC time series in the re-parameterized model (when run without hosing), has almost the same mean as before, but much larger autocorrelation due to the successful destabilization. As expected from theory \citep{WiesenfeldMcNamara1986, Scheffer2009}, and in similarity to the double-well system (Fig. \ref{fig: double well}b) the time series consequently also has larger variance even though the additive state noise was the same.

\section{Analysing the computational cost of the method}\label{sec: Analysing the computational cost of the method}

An important question regarding our calibration method for adjusting stability biases in complex system models is how well it performs in terms of computational cost when compared to a brute force parameter search. In particular, we are interested in how the computational cost scales with the number of parameters in our model.

Given a target stability of the dynamical system in question, the computational cost for finding a suitable parameter combination would scale exponentially with the number of adjustable parameters when applying a brute force parameter search. This manifestation of the so-called curse of dimensionality has so far prevented systematic calibration of climate or Earth system models, also because they are not differentiable. In the case of our method, the number of iterations needed to achieve the destabilization goal does not necessarily show any systematic dependency on the number of parameters, as the parameters evolve along a gradient of decreasing or increasing linear stability of the system. We therefore expect that the significant factor determining how the computational cost scales with the number of parameters will depend on how the length of the trajectory needed to estimate the parameter-dependent VAR model in each iteration step scales with the number of parameters. Since the length of such a trajectory would typically be independent of the number of parameters in a brute force setting, the main question is whether the length of the trajectory in an iteration step scales sub-exponentially. 

We determine the scaling behavior of the trajectory length in the two example systems presented above. To this end, we consider each system with its respective initial parameter constellation. Then, in a first step, we only vary one parameter at a time, fixing all other parameter values. Using a trajectory of fixed length, we collect the errors in the coefficient $A_{op}$ of the resulting VAR model for each separate parameter variation. This results in a vector $\sigma^2_{p_i} \in \mathbb R^{d_p}$, where the $i$-th entry corresponds to the variances of $A_{op}$ in a VAR model where only the $i$-th parameter is varied. The reason why we only focus on the error of $A_{op}$, as opposed to also considering further coefficients, is due to simplicity and the fact that $A_{op}$ is the most relevant coefficient for parameter updates when purely aiming for a change in stability.

In a second step, we iterate through all possible parameter constellations, including at least two parameters. For each such parameter set, we iteratively increase the trajectory length used to compute a VAR model in each iteration, and compare the variance of the sum of all entries of $A_{op}$ (taking correlations into account) of the resulting VAR model to the sum of the respective entries in $\sigma^2_{p_i}$. As soon as the variance of the sum of the entries in $A_{op}$ drops below the latter sum, the current trajectory length necessary to fulfill this condition is noted. The reason why we do not directly compare the variances of each coefficient in $A_{op}$ to their counterpart in $\sigma^2_{p_i}$ is that this can lead to extremely long trajectory lengths necessary to meet this condition, in the case that the entries in $\sigma^2_{p_i}$ differ in magnitude. Thus, if the total number of parameters that can possibly be varied is given by $d_p$ and the currently considered set of parameters consists of $p$ parameters, this results in $\begin{pmatrix}
    d_p \\ p
\end{pmatrix}$ trajectory lengths, corresponding to the case of $p$ parameters being varied. By averaging all values corresponding to a given value of $p$ and visualizing these data, we can try to infer the functional dependence of the trajectory length with respect to the number of parameters that are varied.

For each given parameter set, we always use the same noise amplitudes $\sigma_i$ for a given parameter $p_i$ regardless of the other parameters it is paired with. In the case of the double-well system we used 0.05$\cdot \sqrt{1-\lambda_p^2}$ (see Sect. \ref{sec: Double well system}) as noise amplitude for every parameter. For the five-box AMOC model we used $(\sigma_\gamma, \sigma_\eta, \sigma_{K_N}, \sigma_{K_S}, \sigma_{K_{IP}}) = (0.020, 5.243, 0.328, 0.082, 2.621) \cdot \sqrt{1-\lambda_p^2}$, where we determined the noise amplitudes using the procedure described in Sect. \ref{sec: AMOC 5-box model}. As this choice of noise amplitudes for the five-box model led to quite long trajectory lengths necessary to meet the convergence condition, we also empirically modified the noise amplitudes to be $(\tilde{\sigma}_\gamma, \tilde \sigma_\eta, \tilde \sigma_{K_N}, \tilde \sigma_{K_S}, \tilde \sigma_{K_{IP}}) = (0.020 \cdot 32, 5.243/32, 0.328/2, 0.082 \cdot 2, 2.621/8) \cdot \sqrt{1-\lambda_p^2}$, which resulted in much shorter trajectory lengths. We will refer to the resulting model as the optimized box model system in the following.

%\begin{itemize}
    %\item[1.] 
When determining the values of $\sigma^2_{p_i}$ we used VAR models computed from data series of length 6000 in the case of the double well system and length 1000 for the two versions of the five-box model. In order to suppress stochastic effects, we repeated this procedure 100 times and took the final value of $\sigma^2_{p_i}$ to be the average over all of these runs.
    %\item[2.] 
For each parameter set including at least two parameters, we computed the trajectory length needed to fulfill the condition described above as the average over 20 such runs. To this end, we increased the trajectory length by 1000 in each iteration step, starting from 6000 for the double well system and from 6000 for the optimized five-box model system). For the non-optimized five-box model, we increased the trajectory length in the following fashion: $10^3, 2\cdot 10^3, 5 \cdot 10^3, 10^4, 2\cdot 10^4, 5 \cdot 10^4, 10^5, ...$ to save computation time.
%\end{itemize}

\begin{figure*}
     \centering
     \includegraphics[width=\linewidth]{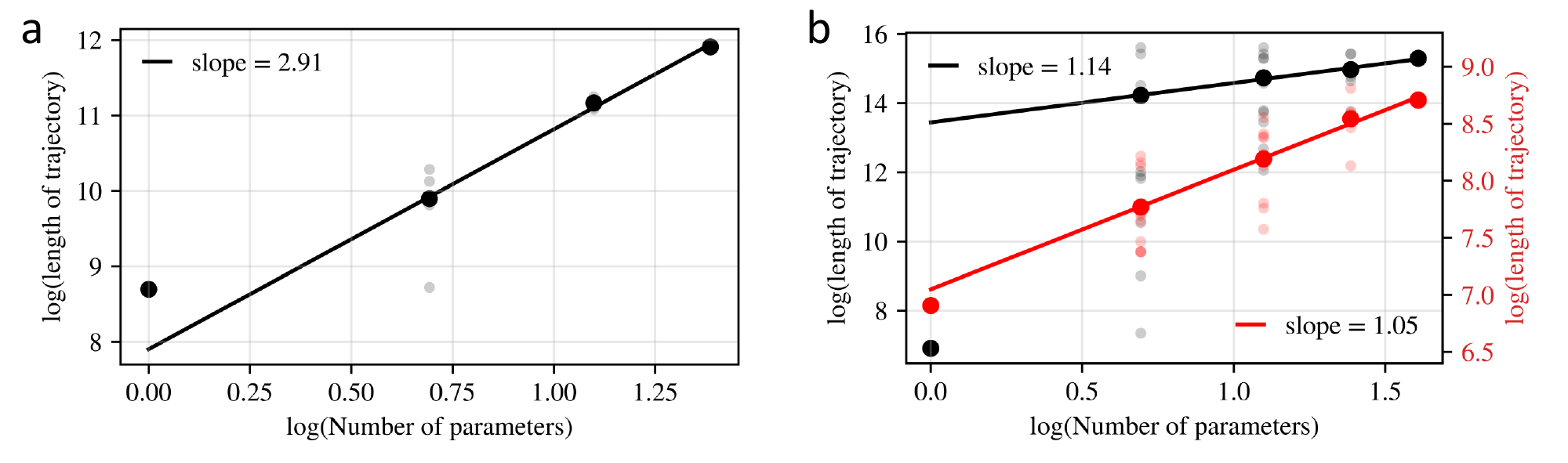}
        \caption{Computational cost of our calibration method to adjust stability for (a) the double well system, (b, black) the box model system, and (b, red) the optimized box model system.
        Both figures show how the length of the trajectory that is needed in a single iteration step in order to fulfill the accuracy condition (see Sect. \ref{sec: Analysing the computational cost of the method}), depends on the number of parameters under consideration.
        The small transparent data points correspond to (averaged) measurements for a given fixed parameter subset whereas the large data points display the average value taken over all parameter subsets, containing the same number of parameters. By fitting a linear function to the averaged data points in the log-log plot for each system respectively, we infer a polynomial dependence of the trajectory length on the number of parameters. We excluded the data points corresponding to one parameter being varied from the linear fits due to its unique significance for the generation process of the shown data.}
        \label{fig: computational cost}
\end{figure*}

The results of the analysis on the computational cost of the method are shown in Fig. \ref{fig: computational cost}. We notice that the data, displayed in a log-log plot, agree well with respective linear fits. One can also observe that the range of trajectory lengths, which we observe for parameter subsets with a fixed number of parameters (small data points), decreases as this number grows in all three cases.

\section{Discussion}\label{sec: Discussion}

In both examples that we considered, namely the simple double-well system and the five-box model of the global ocean circulation, our method is able to adjust the stability of the modeled system. Specifically, focusing on the task of reducing stability, the largest eigenvalue $\lambda$ is moved closer to 1. The computational cost when destabilizing the double well system as well as the five-box model shows a polynomial dependency on the number of parameters under consideration, although the leading order power seems to differ between the two models (Fig. \ref{fig: computational cost}). This polynomial dependency implies great improvement compared to a brute force parameter search (e.g. the exponential scaling when populating the parameter space with always the same density). This is particularly relevant for non-differentiable models, where gradients of model output with respect to parameter variations cannot be computed. However, brute-force approaches are prohibitive due to the enormous computational costs of running these models. Ensemble Kalman Inversion (EKI) \citep{Iglesias2013, Kovachki2019} may be a potential alternative, but several caveats would have to be overcome. For example, (i) the subspace spanned by the initial EKI ensemble must already cover the solution, which can require a very large (initial) ensemble; (ii) though EKI can scale well (i.e. approximately quadratically with the number of parameters), it does not provide any guaranty regarding the convergence time or the quality of the estimated parameters (i.e. closeness to a global or local minimum); (iii) it may be challenging to apply constraints, e.g. conserving the system's mean state during parameter updates, and (iv), since the system's stability is a property of the dynamics rather than an observable, one would need to artificially construct an ``observed'' property (including its uncertainty), which may pose problems. For example, the maximum value of a bounded value, like autocorrelation = 1, cannot be a valid target for EKI.

%Also after finding the correct hyperparameters, i.e. establishing all the system specific sub-processes (e.g. calibration of the noise amplitudes or even choice of suitable observables), we were able to destabilize the systems without having to integrate the system for a disproportionate amount of time steps. 
% local method, not global

We note that our method is not designed to find the most unstable version of the system. Since the result depends on the specific initial conditions of the system state and the parameters, its purpose is to efficiently approach regional minima of the stability landscape. Because there may in principle be multiple unstable regions, the search algorithm may need to be run many times from different initial parameter sets, and the computational efficiency of our method is therefore crucial. In order to scan larger regions in parameter space, global methods have to be used. For example, one can perform perturbed-parameter ensemble simulations \citep{Collins2011}, where the combinations of model parameters are determined by a Latin Hypercube sampling \citep{Urban2010}. Such global approaches and our local optimization method are complementary approaches that can be combined in a novel way to illuminate a model's dependence on parameters. In cases where more physical or observational constraints are applied to the parameter ranges and the resulting equilibrium solutions than we used here, the resulting target region of parameter space may already be rather constrained, and our local search method may indeed already deliver a global optimum by itself.

%% finding suitable hyperparams is hard
A practical challenge associated with applying our method is that it can be difficult and computationally expensive to find suitable hyperparameters, involving the calibration of the amplitudes of the noise imposed on the system state or the parameters, or the choice of suitable observables. In general, the final parameter configuration can potentially depend on the choice of the hyperparameters. Depending on how we choose the observables and the state noise, our method offers flexibility as to which aspects the system is stabilized or destabilized (through the choice of observables) and also what types of perturbation are relevant (by choosing suitable noise).

%%% generalization to other systems
The general recipe provided above has been explicitly demonstrated for the two example systems in Sect. \ref{sec: application}, but it will also work for many other systems. There is no guaranty, however, that the implementation we chose is also the most optimal approach in a specific case with a considerably different model. Depending on the model at hand, users will need to reconsider how to introduce state noise, how to calibrate parameter noise levels, and how to constrain certain parameters when updating them. The small additions and variations applied in our two example systems in this regard showcase how the method may be adapted. Moreover, we also propose a number of additional methods to evaluate the quality of the VAR model fit (Appendix \ref{sec: VAReval}), to test the whiteness of the residuals (Appendix \ref{sec: whitenesstest}), and an approach to improve the VAR estimation by denoising (Appendix \ref{sec: denoising}). We implemented and applied the latter method to our two example systems, which yielded similar results to those discussed above (not shown). 

%% Caveats / further improvements
%% reaching limits of the parameter ranges
The fact that the result depends on the constraints on the parameter values is a desired property. For example, parameters in complex climate models are often constrained by observations or by their physical meaning. Considering the five-box model, we notice that even after 500 iteration steps we were not able to push $\lambda$ as close to 1 (Fig. \ref{fig: 5-boxmodel_destabilization}b) as in the case of the double well system. This is likely due to the fact that some parameters hit the boundaries of their restricted ranges (they become zero, see Fig. \ref{fig: 5-boxmodel_destabilization}a), which might prevent the system from reaching an arbitrary degree of destabilization. In other words, one cannot always expect to be able to change the stability of a system’s fixed point, given a set of changeable parameters. Such a result is informative, since it points to structural limitations of a model (or, potentially, the real world) which cannot be overcome and which prevent tipping behavior. By integrating domain-specific constraints and real-world observables, our method enables the exploration of climate change trajectories that are consistent with observations and theory, yet expose possible worst-case tipping-point scenarios.

%% 5box, convergence
An indicator of potential improvements is the convergence of the tuned parameter values. In general, each parameter has a clear tendency for the system to lose stability, despite the stochastic elements in our approach. We see an exception to this within the first $\sim 50$ iteration steps for the five-box model, where some parameters seem to evolve contrary to their overall tendency (Fig. \ref{fig: 5-boxmodel_destabilization}a). This effect seems to be correlated with the mean value $\bar q$ of $q$ that significantly deviates from its target value. A possible explanation could be that at this point the VAR model does not extrapolate well, causing parameter changes to affect the system in an unforeseen way. On the one hand, this issue could be fixed by temporarily and adaptively decreasing the step size of the parameter updates between iteration steps. On the other hand, this is not required as long as q recovers back to its original value after a temporary anomaly. Moreover, the fixed step size reduces the computational cost of our method.

%Also notice how the optimized choice of noise amplitudes $\sigma_i$ in the box model system has a severe impact on the performance of the model in terms of computational cost compared to the non-optimized version. This stresses the importance of the choice of suitable noise amplitudes.

% DISCUSSION; ideas to extend the method
%Updating the parameters
% This relates to step 5
Our method is flexible in the requirements that one imposes on the updated parameter values, which in general could
\begin{itemize}
    \item preserve the equilibrium values of the observables by fulfilling Eq. \ref{eq: preserving bar o}.
    \item only take on allowed values in case the parameters are restricted.
    \item not exceed a certain step change, which would typically depend on the noise amplitudes $\sigma_i$ used for the parameter variation, as these determine the explored region in parameter space.
    \item minimize the errors of the new value of the maximum eigenvalue or the error on the accuracy on how well the condition for preserving the observable values is fulfilled.
\end{itemize}
All of these conditions (and more) can be included in the parameter update by either implementing these as constraints or as penalty terms in the optimization problem that needs to be solved in order to determine the new parameter values. For example, one could constrain the variance of the observable to some target range (e.g matching an ESM) by adding an additional penalty to Eq. \ref{eq:maximise}.

%Both of the examples (double well and 5-box model) discussed in the previous sections, confirm that the method can work in principle and has the advantage of providing a systematic approach on how to simplify and destabilize complex systems. 

Further research on this topic could be dedicated to applying our method to more complex models and finding more efficient ways for determining appropriate hyperparameter values. Also, a more detailed investigation on why the computational efficiency scales polynomially with the number of parameters, compared to exponential scaling for brute force methods, or more sophisticated methods like EKI, could be of great interest. In addition, machine learning methods \citep{Baumeister2018, Berkenkamp2023, Hourdin2023} could be used, instead of VAR fitting, for targeted calibration to control system stability.

Our implementation of the method is fully automated, i.e. all steps outlined above, including the noise calibration and iterative parameter updates, work without intervention (see Sect. \ref{sec:code} and code documentation online). In principle, it is straightforward to apply our method to more complex models, though the parameter updates and the additive state noise would require one to incorporate the compilation of the model code into the procedure. In Earth system models of intermediate complexity (EMICs), internal dynamics are often missing or already parameterized as noise \citep{Willeit2022}, rendering them particularly suited for our method (which can exploit knowledge on the realization of the imposed state noise). The same applies to complex components of ESM with little internal variability, such as (parts of) dynamic global vegetation models \citep{Bathiany2024} or ice sheet models \citep{Robinson2020}. Relevant problems our method could tackle in EMICs are, for example, (i) the ``Green Sahara'' problem \citep{HopcroftValdes2021}, where parameter combinations controlling precipitation and vegetation dynamics are required that allow reproducing abrupt shifts as seen in paleo reconstructions, and (ii) the potential for Amazon forest dieback via parameters affecting moisture recycling and/or vegetation fire dynamics \citep{Cox2004, LovejoyNobre2018}.

Regarding comprehensive ocean or atmosphere general circulation models, a more fundamental caveat can be that the ``noise level'' of the observables cannot be directly controlled since their variability emerges from chaotic internal dynamics on short timescales. However, such internal variability does not exclude using our method in principle. For example, depending on the observable, the internal variability can still be sufficiently small, which is, in fact, the case when averaging salinities from ocean GCMs to the scale of the five-box model \citep{Chapman2024a}. Even in case of larger variance, the method may still yield useful results, although it will generally become less accurate since second-order terms cannot be estimated well anymore due to lack of knowledge of the ``noise'' realization (see Appendix \ref{sec: step4_secondorderterms}). A possible solution could be to fit a model to the slow internal modes, allowing one to separate its recovery dynamics from the noise term (fast chaotic geophysical fluid dynamics). In this situation, the presence of internal variability would make it unnecessary to add any artificial noise, which would improve the usability of the calibration method.

In the case of analyzing slow Earth system components (coupled to an atmosphere and/or ocean GCM), the most important limitation of our method is probably that the large number of uncertain parameters, and the long timescales of the processes of interest, make direct application of our method to calibrate parameters computationally difficult for some problems (including AMOC stability in ESMs). Although the iterative parameter updates work only sequentially, the efficiency of the method may be improved by using an initial condition ensemble to generate the time series used in step 1 in each iteration.

We see our method as an efficient way to identify the key targets for ESM tuning, to access model versions with realistic stability properties. The right choice of such metrics is not obvious {\it a priori}. Our method is probably best suited to exploring the parameter space of physically-based emulators such as the AMOC box model used in our study, to identify unstable regions of its parameter space that are consistent with available observational constraints. The five-box model can be calibrated to represent different comprehensive ESMs \cite{Wood2019, Chapman2024a, Chapman2024b} and is a good example of how applying our method to a reduced-order model (where technical and theoretical demands for users are still tractable), can be informative for the stability of complex models. Because the parameters of the AMOC box model themselves correspond to emergent (observable, large-scale) properties of the climate system \citep{Wood2019}, identifying the unstable parameter regions provides a framework for understanding why different ESMs have different stability properties. 
A pipeline outlining how our approach can support parameter tuning in a given comprehensive ESM proceeds as follows: 1. Calibrate a physically-based emulator to the ESM. 2. Run a targeted stability calibration in the emulator parameter space. 3. Insert the resulting ``unstable but still observationally consistent'' parameters back into the ESM.

In case the ESM features larger internal variability than the emulator, the identified optimal parameters may already cause a transition to an alternative regime (``N-tipping'', see \citep{Ashwin2012}), but since our method has tracked the parameter updates, one can step back from the brink as far as required. The optimal parameter regions serving as initial parameter conditions for our method become a set of target metrics for ESM tuning, which can be performed either through conventional tuning approaches, or using systematic methods that are now being developed at several major modeling centers (e.g. \cite{Peatier2022}).

%% beyond destabilization
Due to its flexibility regarding the purpose of the model, the nature of the constraints, and the target property to be optimized, our method is not restricted to applications with the goal of varying the stability of climate models, but can potentially be applied to a wide class of optimization problems within the context of complex dynamical systems.

\section{Conclusion}
We have introduced a targeted method for systematic and objective parameter calibration to adjust system stability in non-differentiable complex system models, under given physical and observational constraints. Our method considers dynamics on the combination of a given model's phase and parameter space and exploits the phenomenon of critical slowing down to identify the optimal direction in parameter space to adjust the stability of modeled systems in a desired way. This makes the method computationally highly efficient, breaking the curse of dimensionality by scaling only polynomially in the number of parameters. Our results are particularly promising given the persisting concerns that major Earth system components are too stable in state-of-the-art models, which are very challenging to calibrate objectively.

\subsection*{Code}
\label{sec:code}
The Julia code implementing our method, as well as the two numerical models used in this study, can be found at \url{https://github.com/TUM-PIK-ESM/targeted-calibration} and is published under the MIT license.

\begin{acknowledgments}
This is ClimTip contribution \#4; the ClimTip project has received funding from the European Union's Horizon Europe research and innovation programme under grant agreement No. 101137601. N.B. and S.B. also acknowledge funding by the Volkswagen foundation. We are grateful to Brian Groenke for helpful discussions, to Max Gelbrecht for education and support concerning Julia programming, to Mycroft Holmes for vital clues given from his armchair, and to two anonymous reviewers who helped us to improve the manuscript by their constructive concerns and suggestions.
\end{acknowledgments}

\section{Author contributions}
N.B. and S.B. defined the research goal, D.P. designed the implementation and execution of the method under supervision of N.B. and S.B., D.P. and R.W. implemented the five-box model, D.P. wrote a first description of the methods and results, S.B. wrote the paper with contributions from all authors.

\appendix

\section{VAR(1) model estimation (step 2)} 
\label{sec: step2_VARmodel}
We assume a general setting in which we are given two time series $(\vec z_t)$ and $(\vec y_t)$ with $t \in \{0,...,T\}$, which can potentially be identical. Our goal is to find a model of the form
\begin{align} %\nonumber
    \vec z_{t+1} = \vec \nu + A \vec y_{t} + \vec \varepsilon_t
\end{align}
which fits the data best in the sense of a least-squares estimate concerning the error vectors $\vec \varepsilon_t$. 

In order to provide compact formulas, we introduce the notation
\begin{align}
    Z &= (\vec z_1,...,\vec z_T) \; \; \; && (d_z \times T) %\nonumber 
    \\
    Y_t &= \begin{pmatrix}
        1 \\
        \vec y_t
    \end{pmatrix}
    && ((d_y+1) \times 1) %\nonumber 
    \\
    Y &= (Y_0,...,Y_{T-1}) &&((d_y+1) \times T) %\nonumber 
    \\
    E &= (\vec \varepsilon_1,..., \vec \varepsilon_T) && (d_z \times T) %\nonumber 
    \\
    B &= (\vec \nu, A) && (d_z \times (1 + d_y)) %\nonumber
\end{align}
where $d_z$ and $d_y$ are the dimensions of $\vec z_t$ and $\vec y_t$ respectively. We then find (also see Eq. (3.2.10) in \cite{Luetkepohl}) that the unbiased least square estimate for $B$ is given by
\begin{align} %\nonumber
    B = Z Y^T (Y Y^T)^{-1}
\end{align}

Further, we also obtain an unbiased estimate for the covariance matrix of the error terms $\vec \varepsilon_t$ given by
\begin{align} \label{eq: Sigma_u}
    \Sigma_\varepsilon = \frac{T}{T- d_y - 1} Z \left(\mathbb{I}_T - Y'(Y Y')^{-1} Y \right) Z'
\end{align}
as described in Eq. (3.2.19) in \cite{Luetkepohl}.

In order to give an expression of the errors and correlations of the coefficients in matrix $B$ we introduce the following:
\begin{align} %\nonumber
    \vec \beta = (B_{1,1}, B_{2,1}, ... , B_{d_z,1}, ... , B_{1, 1+d_y}, ... , B_{d_z, 1+d_y})^T
\end{align}
The covariance matrix of $\vec \beta$ (for asymptotically large $T$) can then be estimated as 
\begin{align} \label{eq: Sigma_beta}
    \Sigma_{\beta} = \frac{\left(\frac{Y Y'}{T}\right)^{-1} \otimes \Sigma_\varepsilon}{T}
\end{align}
which can be found in section 3.2.2. of \cite{Luetkepohl}.

We also estimate the errors of the coefficients by using $(\vec z_t) = (\vec o_t)$ and $(\vec y_t) = ((\vec o_t^T, \vec p_t^T, \vec o_t^T \otimes \vec p_t^T)^T)$.
We make use of these errors when calibrating the noise amplitudes to test the dependence of the observables on the varying parameters (see Appendix \ref{sec: step3_noiseamplitudes}).

\section{Finding appropriate noise amplitudes for the parameter variation (step 3)}
\label{sec: step3_noiseamplitudes}
The optimal choice of the noise amplitudes $\sigma_i$ when generating the parameter series as presented in Eq. \ref{eq: parameter variation} can be problem-specific, when finding a satisfactory trade-off between the importance of well chosen noise amplitudes and the time it takes to compute these. Quality features of well chosen noise amplitudes are that the respective coefficients of $A_p$ and $A_{op}$ from Eq. \ref{eq:VARvary} significantly differ from zero while having small errors. In principle, we aim to maximize each noise level in order to maximize the volume in parameter space which informs us on the effect of the parameters. However, the noise amplitude should not be too large to assure that the parameter variations do not overshadow the additive state noise allowing us to also estimate the parameter independent parts of the model properly. Moreover, no potential restrictions on the parameter ranges should be violated during the parameter variation process, and the overall structure of the dynamical system (close to the considered equilibrium) should remain preserved. We consequently implement a method that tries different noise levels iteratively. In general, the starting value should be sufficiently small; in case of prior knowledge about the system, larger initial values make the method more efficient (we hence picked different initial values for both systems in Sect. \ref{sec: application}).

Our general approach is to vary the noise levels one by one, i.e. we perform the iteration for each parameter separately. Varying the parameters all together is, of course, more efficient. However, this only works if all parameters have a similar influence at similar noise amplitudes, which applies to the special case of the double-well system but not to the five-box model (see Sect. \ref{sec: application}). We also note that for most systems it might not be necessary to update the noise amplitudes in each iteration, which can save computational costs (as is the case in the five-box model).

%% What we do in each case:

%DWS: until variance of observable x exceeds 10**-3. Noise levels are all updated simultaneously. Then take the final noise amplitude. We then also test if the param values are different from 0. 

%5box: until variance of observable q exceeds 0.01 (the value it has without param variations). Each parameter is calibrated separately. Then we consider *all* results which fulfill the "causality condition" ?!. Then we pick combinations for which the values of VAR(q) do not differ more than a factor of 4. => There is only this one extra criterion in AMOC model, and the fact that we change parameters separately (which would be the default; less efficient, but more careful).
%%%The default is to vary each parameter separately, though in some systems, efficiency can be gained by varying them together (the double-well system is such a case).

The iteration ends when the variance of an observable substantially exceeds the variance we obtain without any parameter noise (by setting a fixed threshold value somewhat larger than this natural variability). Thereby, the parameter variation does not overshadow the additive noise, allowing us to also estimate the parameter independent parts of the model properly. In addition, we check whether the observables show a relationship to the varying parameter by testing if the VAR model coefficients differ from zero. 

As a more sophisticated approach, one may also test if the respective coefficients of $A_p$ and $A_{op}$ significantly differ from zero when considering a given parameter $p_i$ as shown in section 3.6 in \cite{Luetkepohl}:
Let $C$ be a $(n \times d_z(d_y +1))$ matrix such that $C \vec \beta$ only consists of the $n$ coefficients in $ \vec \beta$ that are relevant for the coupling of the parameter under consideration with $\vec o$. Then we can compute the following statistic
\begin{align} %\nonumber
    \lambda_F = \frac{1}{n} (C {\vec \beta})' \left[C ((YY')^{-1} \otimes \hat{\Sigma}_\varepsilon) C' \right]^{-1} C {\vec \beta}
\end{align}
which we expect to follow an $F(n,T-d_y-1)$-distribution in the case where there is no causal relationship from $\vec p$ to $\vec o$.

In order to find suitable noise amplitudes, possible approaches could also include grid searches or more sophisticated methods where the noise amplitudes are simultaneously varied until certain conditions are met (in similarity to the double-well system, Sect. \ref{sec: Double well system}).

\section{Estimation of VAR model second order terms (step 4)} 
\label{sec: step4_secondorderterms}
If the magnitude of the variations of the observables $\vec o$ and the parameters $\vec p$ are significantly smaller than the equilibrium values, the second-order term, $A_{op}$, is difficult to estimate even from a long time series. This, however, inevitably leads to wrong predictions also of the coefficients describing higher order terms. This can be understood by expressing 
\begin{align}
\vec p_t = \vec p_{init} + \Delta \vec p_t, \\
\vec o_t^{vary} = \vec o_{eq} + \Delta \vec o_t^{vary},
\end{align}
where $\vec o_{eq}$ is the current equilibrium value of the observable for $\vec p = \vec p_{init}$. Rewriting Eq. \ref{eq:VARvary} yields
\begin{eqnarray}
\vec o_{t+1}^{vary} &=& \left(\vec \mu + A_o \vec o_{eq} + A_p \vec p_{init} + A_{op} (\vec o_{eq} \otimes \vec p_{init})\right) \nonumber \\ && + \left(A_o \Delta \vec o_t^{vary} + A_{op} (\Delta \vec o_t^{vary} \otimes \vec p_{init})\right) \nonumber \\ && + \left(A_p \Delta \vec p_t + A_{op} (\vec o_{eq} \otimes \Delta\vec p_t) \right) \nonumber \\ && + \left(A_{op} (\Delta \vec o_t^{vary} \otimes \Delta \vec p_t ) \right) \nonumber \\ && + \vec \varepsilon_t^{vary}.
\end{eqnarray}
$A_{op}$ can only be estimated from the second order term $A_{op} ( \Delta \vec o_t^{vary} \otimes \Delta \vec p_t)$ and, if incorrectly predicted, will distort also the approximation of the higher order coefficients, which leads to wrong conclusions concerning the parameter dependence of the linearized model.

In order to solve this problem, we compare each predicted future time step from the full VAR model to a prediction using a model without parameter dependencies, where we use our knowledge of the state noise trajectory $(\vec u_t^{(x)})$ used to force the dynamical system.
The trajectory of the observable for the system with fixed parameters, $(\vec o_t^{fixed})$, is computed by
    \begin{align} %\nonumber
        \vec o_{t+1}^{fixed} = f_o\left(f_p(\vec x_t^{vary}, \vec p_{init}) + \vec u_t^{(x)} , \vec p_{init}\right)
    \end{align}
Assuming that the time evolution of the observables can indeed be described as suggested in Eq. \ref{eq:VARvary}, we obtain
    \begin{eqnarray}
        \label{eq:VARvarycomplex}
        \vec o_{t+1}^{vary} &=& \vec \mu + A_o \vec o_t^{vary} + A_p \vec p_t \nonumber \\ && + A_{op} \left(\vec o_t^{vary} \otimes \vec p_t \right)  + \vec \varepsilon_t^{vary} %\nonumber
        \\
        \label{eq:VARfixedcomplex}
        \vec o_{t+1}^{fixed} &=& \vec \mu + A_o \vec o_t^{vary} + A_p \vec p_{init} \nonumber \\ && + A_{op} \left(\vec o_t^{vary} \otimes \vec p_{init} \right) + \vec \varepsilon_t^{fixed}
    \end{eqnarray}

where Eq. \ref{eq:VARvarycomplex} is Eq. \ref{eq:VARvary} from above, and is repeated here for practical purposes.
    
Defining the time series $(\hat o_t) \equiv (\vec o_t^{vary}- \vec o_t^{fixed})$, and taking the difference Eq. \ref{eq:VARvarycomplex} - Eq. \ref{eq:VARfixedcomplex}, we find
\begin{align} %\nonumber
        \hat o_{t+1} = A_p \Delta \vec p_t + A_{op} \left( \vec o_t^{vary} \otimes \Delta \vec p_t \right) + \Delta \vec \varepsilon_t 
\end{align}
with $\Delta \vec p_t  \equiv \vec p_t - \vec p_{init}$ and $\Delta \vec \varepsilon_t \equiv \vec \varepsilon_t^{vary} - \vec \varepsilon_t^{fixed}$. If we now fit a VAR(1) model to the data $\vec z_t = \hat o_{t}$ and $Y_t = (\Delta \vec p_t^T, (\vec o_t^{vary} \otimes \Delta \vec p_t)^T )^T$ (see Appendix \ref{sec: step2_VARmodel} for details of the notation) we can estimate $A_p$ and $A_{op}$. Note that by using $Y_t$ instead of $y_t$ we assume $\vec \nu = 0$ in our estimation. By eliminating the zeroth order terms as well as the term of first order in $\Delta \vec o_t^{vary}$ the model becomes much simpler and it is therefore easier to estimate the remaining model-parameters. In particular, there is no longer an ambiguity whether variations in the observable $\vec o_t$ must be captured by the coefficient $A_o$ or by $A_{op}$ making it easier to correctly estimate $A_{op}$, despite the fact that it nevertheless describes a second order contribution.

In order to determine all coefficients from Eq. \ref{eq:VARvary} (Eq. \ref{eq:VARvarycomplex}), we equate the VAR model from Eq. \ref{eq:VARfixed} to the model from Eq. \ref{eq:VARfixedcomplex}, as both models describe the evolution for parameter values fixed to $\vec p_{init}$. By comparing coefficients, we obtain 
\begin{align}
        \vec \mu =& \vec \nu - A_p \vec p_{init} %\nonumber 
        \\
        A_o =& A - A_{op} ( \mathbb{I}_{d_o} \otimes \vec p_{init}) \,. %\nonumber
\end{align}

We here used the Kronecker product properties:
\begin{align}
A_{op}(\vec o \otimes \vec p) = \big(A_{op}( \vec o \otimes \mathbb{I}_{d_p})\big)\,\vec p \nonumber  \\
= \big(A_{op}( \mathbb{I}_{d_o} \otimes \vec p ) \big)\vec o
\label{eq:Kronecker}
\end{align}

$A_p$ and $A_{op}$ are given as estimated in step 4 in Sect. \ref{sec:Recipe} above. The errors of the coefficients, as well as their correlations, can be computed using error propagation techniques. We consider the errors of both VAR estimations (i.e. the estimation for fixed and for varying parameters) as independent, since they arise from independent data.

\section{Parameter updating under constraints (step 5)}
\label{sec: step5_parameterupdates}

Our aim is to maximize the largest eigenvalue of the VAR model's autoregressive term $A_o + A_{op} (\mathbb{I}_{d_o} \otimes \vec p)$ via parameter changes (step 5 of our recipe in Sect. \ref{sec:Recipe}). We also wish to preserve the mean state (Eq. \ref{eq: preserving bar o}, duplicated here):
\begin{align}\label{eq: preserving bar o duplicated}
        \bar o \overset{!}{=} \vec \mu + A_o \bar o + A_p \vec p + A_{op} (\bar o \otimes \vec p)
\end{align}

Rearranging this equation gives
\begin{align}
\label{eq:constraint_rearranged}
        ( \mathbb{I}_{d_o} - A_o ) \bar o - \vec \mu = (A_p + A_{op}( \bar o \otimes \mathbb{I}_{d_p})) \vec p
\end{align}

where we again used expression \ref{eq:Kronecker}, this time in the form where $\vec p$ stands on the very right, after the brackets.

We abbreviate the left hand side of Eq.  \ref{eq:constraint_rearranged} as $\vec s$, and the term in the brackets on the right hand side as a matrix $n$ (in $\mathbb R^{d_o \times d_p}$):
\begin{align}
    n \equiv& A_p + A_{op}(\bar o \otimes \mathbb{I}_{d_p} )
\end{align}

The constraint \ref{eq: preserving bar o duplicated} implies that any parameter vector $\vec p$ must point on a hyperplane defined by
\begin{align}
    \mathcal{H} = \left\{\vec p \in \ \mathbb R^{d_p}| n   \vec p = \vec s\right\}\,.
\end{align}

The hyperplane consists of all possible parameter combinations that comply with the constraint of keeping the time mean observable fixed. Each row of matrix $n$ is a vector that is perpendicular to this hyperplane, and takes care of a specific component of the observable $\vec o$. 

In the following, we discuss the case of a one-dimensional observable ($d_o=1$), which also applies to the systems we use in Sect. \ref{sec: application}. In this situation, $n$ is a vector of length $d_p$, and s is a scalar, and the eigenvalue to be maximized is the scalar $A_o + A_{op} \vec p$. $A_o$ is constant under noisy parameter fluctuations, and only changes in the next iteration after updating the parameters and fitting a new VAR model. Moreover, the new parameters we seek can be written as
\begin{align}
    \vec p = \vec p_{init} + \Delta \vec p.
\end{align}
where $\vec p_{init}$ is already given.

Hence, the property we optimize is
\begin{align}
    A_{op} \cdot \Delta \vec p %\nonumber 
\end{align}

Here, $A_{op}$ is a vector, and we abbreviate it as $\vec v$.
%(in higher dimensions with $d_o>1$, $\vec v \equiv \mathrm{vec}(A_{op})$, i.e. all observables become stacked into one dimension).
%% No! The second index has do*dp entries, and will not match vector Delta_p with only dp entries!

The condition of keeping the equilibrium value of $q$ at $\bar q$ can now be written as
\begin{align}%\nonumber 
    u \overset{!}{=} \vec n \cdot \Delta \vec p
\end{align}
with
\begin{align}
    u = \bar o (1- A_o) - \mu -  \vec n \cdot \vec p_{init} %\nonumber 
\end{align}

Overall, we hence compute $\Delta \vec p$ by maximizing
\begin{eqnarray}
M &\equiv& \vec v \cdot \Delta \vec p \nonumber
    \\
    &&- \eta_1 (\vec n \cdot \Delta \vec p -u)^2 \nonumber
    \\
    &&- \eta_2 (\Delta \vec p^T, -1, 0,...,0) \Sigma_{nuv} \begin{pmatrix}
        \Delta \vec p \\
        -1\\
        0\\
        \vdots \\
        0
    \end{pmatrix} \nonumber
    \\
    &&- \eta_3 (0,...,0, \Delta \vec p^T) \Sigma_{nuv} \begin{pmatrix}
        0\\
        \vdots\\
        0\\
        \Delta \vec p
    \end{pmatrix}
\label{eq:maximise}
\end{eqnarray}

Here, $\Sigma_{nuv}$ denotes the correlation matrix of $(\vec n^T, u, \vec v^T)^T$ which can be computed from $\Sigma_\beta$ (Eq. \ref{eq: Sigma_beta}): Since $\Sigma_\beta$ contains all errors and their correlations of the VAR model parameters, and since $n$, $v$ and $u$ are functions of these parameters, we can compute their errors and correlations via standard error propagation techniques.

The first penalty term hence penalizes the distance of $\Delta \vec p$ to the hyperplane determined by $\vec n$ and $u$ (to keep the mean of the observable constant).

The terms involving $\eta_2$ and $\eta_3$ in Eq. \ref{eq:maximise} penalize the uncertainties on how $\vec v \cdot \Delta \vec p$ (destabilization of the system) and the first penalty term ($\eta_1$ term) change for a parameter update by $\Delta \vec p$. The second term penalizes the error on $\vec n \cdot \Delta \vec p - u$, and the third term penalizes the error on $\vec v \cdot \Delta \vec p$. Thus, these last two terms in Eq. \ref{eq:maximise} assure that, with a high probability, our parameter update indeed has the desired effect of stabilizing or destabilizing the system and preserving the AMOC strength. The trade-off for this is that the destabilization process could potentially become less efficient. 

Apart from the direction of vector $\Delta \vec p$ as discussed above, we also need to control its length. To this end, we rescale all parameters by their respective exploration range in order to make them comparable, and then demand that the Euclidean norm of this rescaled parameter vector must equal a prescribed value $l$ with

\begin{align} %\nonumber
    l^2 \overset{!}{=} \sum_i \frac{\Delta p_i^2}{l_i^2}
\end{align}
where each $l_i$ measures how large the explored region is for parameter $p_i$. We here use step sizes that match the region in parameter space that we explored via the parameter noise:

\begin{align} %\nonumber 
    l_i = \frac{\sigma_i}{\sqrt{1-\lambda_p^2}}
\label{eq:metric_l}
\end{align}
with $\sigma_i$ and $\lambda_p$ from Eq. \ref{eq: parameter variation}. Additional constraints on certain parameters can easily be implemented (e.g. see Sect. \ref{sec: AMOC 5-box model}).

\section{The five-box AMOC model}
\label{sec: five-box model description}

The model represents the meridional global ocean circulation (``global conveyor belt'') in the form of five boxes. The model has been designed to represent the Atlantic Meridional Overturning Circulation (AMOC), and its connected major circulation features on the globe \citep{Wood2019, Alkhayuon2019}. The set of equations we use is:

\begin{widetext}
\begin{align}
    q =& \frac{\lambda}{1+\lambda \alpha \mu} \left[\alpha(T_S-T_0)+\beta(S_N-S_S) \right]
    \\
    \text{For $q \geq 0$:}
    \\
    V_N \frac{d S_N}{dt} =& q(S_T-S_N)+K_N(S_T-S_N)  + F_N S_0   - F_N S_N  %\nonumber
    \\
    V_T \frac{d S_T}{dt} =& q[\gamma S_S+(1-\gamma)S_{IP}-S_T]+K_S(S_S-S_T)+K_N(S_N-S_T) \nonumber 
    \\
    &  + F_T S_0   +F_N(\gamma S_S+(1-\gamma)S_{IP})+F_S S_S+F_{IP} S_{IP} %\nonumber 
    \\
    V_S \frac{d S_S}{dt} =& q \gamma (S_B-S_S)+K_{IP}(S_{IP}-S_S)+K_S(S_T-S_S)+\eta (S_B-S_S) \nonumber 
    \\
    & +F_S S_0   +\gamma F_N(S_B-S_S)-F_S S_S %\nonumber 
    \\
    V_{IP} \frac{d S_{IP}}{dt} =& q(1-\gamma)(S_B-S_{IP})+K_{IP}(S_S-S_{IP})  + F_{IP} S_0 \nonumber 
    \\
    &  +(1-\gamma)F_N(S_B-S_{IP})-F_{IP} S_{IP} %\nonumber 
    \\
    V_B \frac{d S_B}{dt} =& q(S_N-S_B)+\eta (S_S-S_B)  +F_N(S_N-S_B) %\nonumber
\\
\text{and for $q < 0$:}
\\
    V_N \frac{d S_N}{dt} =& |q|(S_B-S_N)+K_N(S_T-S_N)  + F_N S_0   +(F_T+F_S+F_{IP})S_B %\nonumber
    \\
    V_T \frac{d S_T}{dt} =&  |q|(S_N-S_T)+K_S(S_S-S_T)+K_N(S_N-S_T)  + F_T S_0   -F_T S_T %\nonumber
    \\
    V_S \frac{d S_S}{dt} =&  |q| \gamma (S_T-S_S)+K_{IP}(S_{IP}-S_S)+K_S(S_T-S_S)+\eta (S_B-S_S) \nonumber 
    \\
    & + F_S S_0  +
 \gamma F_T(S_T-S_S)-F_SS_S %\nonumber
 \\
    V_{IP} \frac{d S_{IP}}{dt} =&  |q|(1-\gamma)(S_T-S_{IP})+K_{IP}(S_S-S_{IP})  + F_{IP}S_0 \nonumber 
    \\
    &  +(1-\gamma)F_T(S_T-S_{IP})-F_{IP}S_{IP} %\nonumber
    \\
    V_B \frac{d S_B}{dt} =& |q| \gamma S_S+(1-\gamma) |q| S_{IP}-|q| S_B+ \eta (S_S-S_B)   +F_S(S_S-S_B) \nonumber 
    \\
    &  +F_T[\gamma S_S+(1-\gamma) S_{IP}-S_B]+F_{IP}(S_{IP}-S_B) %\nonumber
\end{align}
\end{widetext}

In contrast to the original model, we do not only demand conservation of salt, but also a conservation of total water volume in each box. This results in two slight differences to the original model:

1. We removed a factor of $\gamma$ from the second-to-last term of Eq. 11 in \cite{Wood2019}.

2. $F_i$ describes the flux of freshwater that the ocean surface of box $i$ exchanges with the atmosphere. In the original model, there is a water flux into the boxes with index $i \in \{N,S,T,IP\}$ if $F_i < 0$, which is not removed from the box, while boxes with $F_i > 0$ lose water volume over time. 

We therefore added additional fluxes between the boxes. The flux between the boxes $N$ and $T$ remains unchanged as this flux is supposed to describe the strength of the AMOC. Furthermore, we split up the additional flux going from box $B$ into the boxes $S$ and $IP$ by the same factor of $\gamma$ as used for the AMOC. 

We use the same parameter values as given in \cite{Wood2019} for the FAMOUS$_A$ simulation, with the exception that in the modified version we use $S_0 = 0.035 psu$ instead of $S_0 = 35$ psu. The reason is that $S_0$ in the modified model can be interpreted as fresh water salinity (the salinity of the rain) whereas in the original model it represents the average salt water salinity. The original equations are then an approximation of our modified equations. As the modification is very small numerically, all results in this study are very likely to be independent of the model version applied.

\section{Evaluation of the VAR(1) model}
\label{sec: VAReval}
As an additional suggestion, we propose to evaluate the accuracy of the VAR approximation by comparing forecasts from the original model and the approximated VAR model, defining some expected error bounds.
Assume that we have an estimated VAR model and $(\vec z_t^{new})$ and $(\vec y_t^{new})$ represent some additional data not used to estimate the model. Then the one-step ahead prediction $\hat z_t^{new}(1)$ of $\vec z_{t+1}^{new}$ is given by $\vec \nu + A \vec y_t^{new}$. As similarly presented in section 3.5 in \cite{Luetkepohl}, the covariance matrix of this one-step prediction is given by
\begin{align} %\nonumber
    \Sigma_{z}(1) = \frac{T + d_y +1}{T} \, \Sigma_\varepsilon
\end{align}
Here, $\Sigma_\varepsilon$ is the covariance matrix of the offsets $\vec \varepsilon$, which can be estimated as given in Eq. \ref{eq: VARmodel_5box}, and $T$ is the length of the data series used to estimate the model.
    
A possible procedure to test the model using one-step forecasts could be the following: Assume we have $n$ validation samples available. Then for each $i \in \{1,...,n\}$ the one step forecast $\vec z_{i}^{new}(1)$ is computed as
\begin{align} %\nonumber
    \hat{z}_{i}^{new}(1) = B Y_{i}^{new} = \vec \nu + A \vec y_i^{new}
\end{align}
We can then use this for each $i$
\begin{align} %\nonumber
    (z_{i}^{new}-\hat{z}_{i-1}^{new}(1))' \, \Sigma_{{z}}(1)^{-1} \, (z_{i}^{new}-\hat{z}_{i-1}^{new}(1)) \thicksim \chi^2(d_z)
\end{align}
If we now pick a random subset $A \subsetneq \{1,...,n\}$ of cardinality $|A| \ll n$ and for $n$ sufficiently large, we can assume that the corresponding one-step forecasts are uncorrelated. This leads to
\begin{eqnarray}
        \sum_{i \in A} (z_{i}^{new}-\hat{z}_{i-1}^{new}(1))' &\,& \Sigma_{{z}}(1)^{-1} \, (z_{i}^{new}-\hat{z}_{i-1}^{new}(1)) \nonumber \\ && \thicksim \chi^2(d_z |A|)\,
\end{eqnarray} %\nonumber
which is a criterion that can be tested.

%% NOT USED!
\section{Testing for whiteness of the VAR residuals}
\label{sec: whitenesstest}
Here we suggest a hypothesis test for quantifying the whiteness of the residuals (see section 4.4 in \cite{Luetkepohl}. This means testing whether the autocorrelation of the series $(u_t)$ vanishes. We set up a model
\begin{align} %\nonumber
    \varepsilon_t = D_1 \varepsilon_{t-1} + ... + D_h \varepsilon_{t-h} + e_t
\end{align}
with $e_t$ as error terms and test the hypothesis $H_0: \; D_1 = ... = D_h = 0$ against $H_1: \; D_j \neq 0$ for at least one $j \in \{1,...,h\}$. 

%\paragraph{Lagrange Multiplier Test:}
Using a Lagrange Multiplier Test, we first define
\begin{align}
    \hat{E} &= Z - B Y %\nonumber 
    \\
    F_i &= \begin{bmatrix}
        0_{(i \times T-i)} & 0_{(i \times i)} \\
        \mathbb{I}_{T-i} & 0_{(T-i \times i)}
    \end{bmatrix} %\nonumber 
    \\
    F &= (F_1,...,F_h) %\nonumber 
    \\
    \hat{\mathcal{E}} &= (\mathbb{I}_h \otimes \hat{E}) F' %\nonumber
    \\
    e &= (e_1,...,e_T) %\nonumber
    \\
    D &= (D_1,...,D_h) %\nonumber
\end{align}
using the same notation as introduced in Appendix \ref{sec: step2_VARmodel}. As described in Appendix C.7 and section 4.4.4 in \cite{Luetkepohl}, it can be shown that the hypothesis test as described above is equivalent to testing whether
\begin{eqnarray} 
    \lambda_{LM}(h) &=& \mathrm{vec}(\hat{E} \hat{\mathcal{E}}')' \nonumber \\ && \left(\left[\hat{\mathcal{E}} \hat{\mathcal{E}}' - \hat{\mathcal{E}} Y' (Y Y')^{-1} Y \hat{\mathcal{E}}' \right]^{-1} \otimes \hat{\Sigma}_\varepsilon^{-1} \right)  \nonumber \\ &&  \mathrm{vec}(\hat{E} \hat{\mathcal{E}}')
\end{eqnarray}
is compatible with a $\chi^2(h d_x^2)$ distribution. To this end, we use that we know the mean value and variance of the $\chi^2$ distribution, which allows us to define a confidence interval where we can test whether or not $\lambda_{LM}$ lies inside it or not.

\section{Improvement of VAR estimation by denoising}
\label{sec: denoising}

%%% How the method could be improved:
%%Removing the artificial noise
In order to potentially increase the accuracy of the VAR estimation with fixed parameters, which is also needed for fixed parameter values (step 2 of our recipe above), one can again exploit the fact that we know the noise values $\vec u_t^{(x)}$ used as offsets for computing the trajectories of the system variables $\vec x_t$.
To this end, we first record the trajectory (including noise) of system variables $(\vec x_t)$, which can then be used to compute the corresponding observable trajectory $(\vec o_t)$ using $f_o$. We then compute a second trajectory $(\tilde o_t)$ defined by 
\begin{align} %\nonumber
    \tilde o_t = f_o(f_p(\vec x_{t-1},\vec p_{init}), \vec p_{init})\,,
\end{align}
i.e. we use the same trajectory but remove the noise in each step. Note that we are considering the case of fixed parameters, which is the reason why we use $\vec p = \vec p_{init}$ as second argument for $f_o$ and $f_p$. If we now estimate the VAR model with $\vec z_t = \tilde o_t$ and $\vec y_t = \vec o_t$ using the procedure and the notation from Appendix \ref{sec: step2_VARmodel}, we maintain the benefits of forcing our system with additive white noise -- i.e. forcing the system out of equilibrium and ``coarse graining'' the Jacobian at the equilibrium to a desired scale -- but remove the noise from the VAR model estimation.
We note that this does not always increase the quality of the VAR model estimation even in the case of simple linear dynamical systems, since $f_o(\cdot , \vec p)$ is not injective in general. This means that multiple configurations of the system variables could lead to the same value for the observable, but the values of the observable in the next time step might differ. In cases where this effect plays a dominant role, the benefit of cutting out the noise forcing might be negligible.
In the case of the five-box model, we found that including the noise correction for the fixed-parameter VAR model only has a negligible effect on the quality of the VAR model estimation, so we did not apply it for simplicity and for the sake of cutting down the computational costs.
%\nocite{*}

% `` ''

\bibliography{sn-bibliography}% Produces the bibliography via BibTeX.

\end{document}